\documentclass[11pt,a4paper]{article}
\pdfoutput=1 
\usepackage{jcappub}
\usepackage{capt-of}
\usepackage{color}
\usepackage[utf8]{inputenc}
\usepackage{amsfonts}
\usepackage{lmodern}
\usepackage{amsthm,graphicx}
\usepackage{epsfig}
\usepackage[normalem]{ulem}
\usepackage{latexsym, amssymb} 
\usepackage{amsmath}
\usepackage{multirow} 
\usepackage{booktabs} 
\usepackage{filecontents}

\usepackage{arydshln} 

\theoremstyle{plain}

\theoremstyle{definition}

\def\be{\begin{equation}}
\def\ee{\end{equation}}
\def\ba{\begin{eqnarray}}
\def\ea{\end{eqnarray}}
\newcommand{\fr}[2]{\frac{#1}{#2}}

\def\m{\text{m}}

\def\P{\text{P}}
\def\FoG{\text{FoG}}

\def\lin{\text{lin}}
\def\D{\text{D}}
\def\N{\text{N}}
\def\E{\text{E}}

\def\R{\text{R}}
\def\A{{\cal A}}
\def\B{{\cal B}}

\def\Gau{\text{Gau}}

\def\Kai{\text{Kai}}

\def\TNS{\text{TNS}}
\def\Quas{\text{Quas}}

\def\fid{\text{fid}}

\def\eff{\text{eff}}
\def\Mpch{\text{h/Mpc}}
\def\Erf{\text{Erf}}
\def\DR12{\text{DR12}}

\title{Reconstruction of real-space linear matter power spectrum from multipoles of BOSS DR12 results}

\author{Seokcheon Lee}

\affiliation{Research institute of natural science, Gyeongsang national university, 501 Jinju Daero, Jinju city, 52828, Korea}

\emailAdd{skylee@kias.re.kr}

\abstract{
Recently, the power spectrum (PS) multipoles using the Baryon Oscillation Spectroscopic Survey (BOSS) Data Release 12 (DR12) sample are analyzed \cite{160703150}. The based model for the analysis is the so-called TNS quasi-linear model and the analysis provides the multipoles up to the hexadecapole \cite{TNS}. Thus, one might be able to recover the real-space linear matter PS by using the combinations of multipoles to investigate the cosmology \cite{0407214}. We provide the analytic form of the ratio of quadrupole (hexadecapole) to monopole moments of the quasi-linear PS including the Fingers-of-God (FoG) effect to recover the real-space PS in the linear regime. One expects that observed values of the ratios of multipoles should be consistent with those of the linear theory at large scales. Thus, we compare the ratios of multipoles of the linear theory, including the FoG effect with the measured values. From these, we recover the linear matter power spectra in real-space. These recovered power spectra are consistent with the linear matter power spectra.} 

\begin{document}



\maketitle

\section{Introduction}

The local concentrations of matter exert a gravitational force on their surroundings, resulting in deviations from the Hubble flow. These peculiar velocities of galaxies contaminate the observed redshifts. This causes the difference in the radial position if the redshift is taken as an indicator of the distance. The spatial distribution of galaxies appears squashed and distorted when their positions are plotted in the redshift-space rather than in the real-space. These effects are called as the redshift-space distortions (RSDs). In the nonlinear regime, the random peculiar velocities of galaxies bound in clusters through the virial theorem cause a Doppler shift to make galaxy distribution elongated toward the observer. This is the so-called the Fingers-of-God (FoG) effect \cite{Jackson, 9603031}. Thus, the power spectrum parallel to the line of sight (l.o.s.) is suppressed. In the linear regime, the peculiar velocities cause an originally spherical distribution of galaxies to look flattened along the line of sight due to its coherent infall. This phenomenon, known as the Kaiser effect, enhances the parallel to the l.o.s. component of the power spectrum \cite{Kaiser}.   

The measured quantities are galaxy redshifts and angles from which one transforms our observables redshift into physical coordinates by assuming specific relations between the redshift and the l.o.s. distance and between the angular separation and the distance perpendicular to the line-of-sight given by the fiducial cosmological model. The comoving radial directional distortions depend on the Hubble parameter. And the angular distortions ({\it transverse distance direction}) are presented in the angular diameter distance. If the fiducial cosmological model is different from the true cosmology, then it will produce geometric warping and artificially introduce the anisotropic distortions independently from the effect of the redshift-space distortions. These distortions are called as the Alcock-Pacyznski (AP) effects \cite{AP}. In addition to RSDs, the AP effects provide the cosmological information from galaxy surveys. These degeneracies on the anisotropies in the galaxy PS can be broken by using the Baryon Acoustic Oscillations (BAO) feature in the PS \cite{9605017, 12036594, 12065309, 160703153}. 

The analysis of the galaxy multipole power spectrum is used to measure the growth rate of density perturbations, $f = d \ln \D_{1}/d \ln a$, where $\D_{1}$ is the growth factor. This is used to test General Relativity. The redshift-space galaxy PS including both the RSDs and the scaling factors accounting for the AP effects can be modeled as $\P_{g}(f,b,\sigma,\mu,k,z) = \Bigl(\fr{r_{s}^{\fid}}{r_{s}}\Bigr)^3 \fr{1}{\alpha_{\perp}^2 \alpha_{\parallel}}  \D_{\FoG}^{2}(f,\sigma,\mu,k,z) \Bigl(b(k,z) + f(z) \mu^2 \Bigr)^2 \P_{\text{m}}(k) $, where $\alpha_{\parallel} = \fr{H^{\fid}(z) r_{s}^{\fid}(z_{d})}{H(z) r_{s}(z_{d})}$, 
$\alpha_{\perp} = \fr{\D_{A}(z) r_{s}^{\fid}(z_{d})}{\D_{A}^{\fid}(z) r_{s}(z_{d})}$ where $H^{\fid}(z)$ and $\D_{A}^{\fid}(z)$ are the fiducial values for the Hubble parameter and the angular diameter distance at the effective redshifts of the dataset, $r_{s}^{\fid}(z_{d})$ is the fiducial value of the sound horizon scale at the drag epoch assumed in the power spectrum template, $\mu$ is the directional cosine between the l.o.s. direction and the wave number vector, $b(k,z)$ is the bias factor, $\P_{\text{m}}$ is the linear matter power spectrum, $\D_{\FoG}(k,\mu)$ describes the damping factor due to the FoG effect. The RSDs causes the anisotropy of the clustering amplitude depending on $\mu$. The multipole power spectra are defined by the coefficients of the multipole expansion $\P(k,\mu) = \sum_{l=0,2,\cdots} \P_{l}(k) {\cal L}_{l}(\mu) (2l+1)$, where ${\cal L}_{l}(\mu)$ are the Legendre polynomials. The monopole, $\P_{0}$ represents the angular averaged power spectrum and is usually what we mean by the power spectrum. $\P_{2}(k)$ is the quadrupole spectrum, which gives the leading anisotropic contribution and can be used to constrain the dark energy \cite{0409207}. It is also emphasized that combining the monopole and quadrupole power spectra breaks the degeneracies between the multiple bias parameters and dark energy for both liner and nonlinear regimes. 

The anisotropic clustering of the Baryon Oscillation Spectroscopic Survey (BOSS) CMASS both Data Release 11 (DR11) and DR12 have been investigated in Fourier space, using the PS multipoles to measure cosmological parameters \cite{13124611,160703150}. The analysis demonstrates that the low redshifts ($z_{\eff} = 0.38$ and 0.51) of DR12 constraint are consistent with those of Planck \cite{150201589}. We testify that this statement is correct for the monopole and quadrupole but not for the hexadecapole. Also $k \leq 0.03 \Mpch$ deviates from the theory. It is known that one can recover the real-space power spectrum from the measurements of the redshift-space multipoles by the same linear combination as in the Kaiser limit \cite{0407214}. However, if one includes the FoG damping factor, then one needs to generalize the simple Kaiser limit formula. We provide this formula including FoG effect both for the linear theory and the quasi-linear one.  

Important point is that one might be able to recover the real-space linear matter PS from the quasi-linear redshift-space multipoles as done in the Kaiser limit. From the real-space linear matter PS, the information on the primordial power spectrum, such as the spectral tilt and the shape parameter, can be obtained. Therefore, it is important to check whether the obtained multipoles is accurate enough to reconstruct the real-space PS. The ratios of multipoles can be used for this purpose. We use both the Kaiser limit and the FoG effects to obtain both the theoretical values of multipole ratios and observational ones. We obtain the analytic forms of the ratios of the multipoles for both cases and compare them with the data. Reconstructing the linear matter power spectrum by adopting the ratios of multipoles reduces the errors in the nonlinear effect on the Gaussian FoG effects factor. Thus, this method improves the previous one, which uses the linear combination of multipoles. Also, TNS model includes seven parameters for each redshift bin in each galactic cap. In our method, we can reduce these numbers to three, $P_{\delta\delta}$, ${\cal A}$, and {\cal B}. 

In the next section \ref{LT}, we review the ratios of the multipoles for the Kaiser limit. We also provide the analytic formulae of the linear theory multipole ratios including the FoG effects. We generalize the ratios of the multipoles for the quasi-linear PS in section \ref{QLS}. We also provide the analytic formulae of the multipoles using the quasi-linear PS. The linear matter spectrum in real-space is recovered in the section \ref{RSMPS} We compare the observational data with the linear theory prediction to investigate the quality of observed multipoles. We conclude in section \ref{Con}. The covariance matrices of the ratios of the multipoles for the different modes at each redshift and their chi-square values are given in the appendix.        

\section{Linear Theory}
\label{LT} 
In this section, we review the multipoles of the linear redshift-space power spectrum, both without and with the FoG factor. We also obtain the analytic forms of multipoles and their ratios. The results are independent of the linear power spectrum. We use these formulae to obtain the theoretical values of the multipole ratios of the linear PS.   

The redshift-space position ${\bf s}$ of a galaxy differs from its real-space position ${\bf r}$ due to its peculiar velocity
\be {\bf s} = {\bf r} + v_{z}({\bf r}) \hat{{\bf z}} \label{s} \, , \ee
where $v_{z}({\bf r}) \equiv \sigma_{z}({\bf r})/(aH)$ is the l.o.s component of the galaxy velocity. The galaxy over-density field in redshift-space can be obtained by imposing mass conservation, $(1+\delta_{g}^{s}) d^3s = (1+\delta_{g}) d^3r$. The exact Jacobian for the real-space to the redshift-space transformation is
\be \fr{(1+\delta_{g}) }{(1+\delta_{g}^{s}) } = \fr{d^3s}{d^3r} = \Bigl(1+\fr{v_{z}}{z}\Bigr)^2 \Bigl(1+\fr{dv_{z}}{dz}\Bigr) \, . \label{Jacobian} \ee
In the limit where one is looking at scales much smaller than the mean distance to the pair, $v_{z}/z \ll 1$ it is only the second term that is important
\be (1+\delta_{g}^{s})  = (1+\delta_{g}) \fr{d^3r}{d^3s } \simeq (1+\delta_{g}) \Bigl(1+\fr{dv_{z}}{dz}\Bigr)^{-1} \, . \label{deltags} \ee
If one assumes an irrotational velocity field, then one can write $v_{z} = -\partial/\partial z \nabla^{-2} \theta$, where $\theta = - \nabla \cdot {\bf v}$. In the Fourier space, $(\partial/\partial z)^{2} \nabla^{-2} = (k_{z}/k)^2 = \mu^2$ where $\mu$ is the cosine of the angle between the wave vector, $k$ and the l.o.s. Thus, Eq.(\ref{deltags}) becomes
\be \delta_{g}^{s}(k) = \delta_{g}(k) + \mu^2 \theta(k) \label{deltags2} \, , \ee
to the linear order. Often it is further assumed that the velocity field comes from the linear perturbation theory of the matter
\be \theta(k) = f \delta_{\text{m}}(k) \label{theta}  \, ,\ee
where $f \equiv d \ln \D_{1}/d \ln a$ is the growth index and $\D_{1}$ is the growth factor.  On large scales, the galaxies along the l.o.s appear closer to each other than their actual distance because mass flows from low-density regions onto high density sheets (Kaiser effect) \cite{Kaiser}. 
One can relate the galaxy density perturbation with the matter density one as $\delta_{g} = b(z,k) \delta_{m}$ with $b$ being the so-called bias factor. In the Kaiser limit, the linear redshift-space galaxy power spectrum is given by \cite{11054165}
\be \P_{g} (f, b, \mu, k, z) = \Bigl( b(k,z)+f(z)\mu^2 \Bigr)^2 \P_{\delta\delta}(k, z) \simeq b(z)^2 \Bigl(1+\beta(z) \mu^2 \Bigr)^2 \P_{\delta\delta}(k,z) \label{PgKaiser} \, ,\ee
where $\P_{\delta\delta}$ is the linear matter power spectrum obtained from the linear theory. The assumption that the bias factor is scale independent at the linear regime is used in the second equality.
From the above Eq.(\ref{PgKaiser}), the Kaiser limit linear theory multipoles are given by
{\fontsize 80 
\ba \P_{0}^{\Kai} &=& \frac{b^2}{15} \Bigl(15+10\beta+3 \beta^2  \Bigr) \P_{\delta\delta} = \fr{b^2}{15} \Bigl(15 +10\frac{f}{b} +3 \frac{f^2}{b^2} \Bigr) \P_{\delta\delta} \label{P0Kai} \, , \\ 
\P_{2}^{\Kai} &=& \frac{4}{21} b^2 \beta  (7+3 \beta ) \P_{\delta\delta} = \frac{4}{21} f (7 b+3 f) \P_{\delta\delta} \label{P2Kai} \, , \\
\P_{4}^{\Kai} &=& \frac{8}{35} b^2 \beta ^2 \P_{\delta\delta} =  \fr{8}{35} f^2 \P_{\delta\delta} \label{P4Kai} \, . \ea }
The ratios of these multipoles are well known as 
{\fontsize 80
\ba \R_{2}^{\Kai} &\equiv& \fr{\P_{2}^{\Kai}}{\P_{0}^{\Kai}} = \frac{20 \beta  (7+3 \beta )}{7 (15+ 10 \beta+3 \beta^{2} )} = \frac{20 f (7 b+3 f)}{7 \left(15 b^2+10 b f+3 f^2\right)} \label{R2Kai} \, , \\
\R_{4}^{\Kai} &\equiv& \fr{\P_{4}^{\Kai}}{\P_{0}^{\Kai}} = \frac{24 \beta ^2}{7 (15+10 \beta+3 \beta^{2})} = \frac{24 f^2}{7 (15 b^2+10 b f+3 f^2)} \label{R4Kai} \, .\ea }
These ratios of multipoles are independent of the linear power spectrum, $\P_{\delta\delta}$ and thus one can obtain the growth index, $f$ and the bias factor, $b$ from the observed values of multipoles. The linear matter PS is recovered by using the linear combination of multipoles
\be \P_{\delta\delta} = \fr{1}{b^2} \Bigl( \P_{0}^{\Kai} - \fr{1}{2} \P_{2}^{\Kai} + \fr{3}{8} \P_{4}^{\Kai} \Bigr) \label{PddKai} \, .\ee  

One can extend the above consideration by including the FoG effect. Then the linear redshift-space PS is given by 
\be \text{P}_{g} (f, b, \sigma, \mu, k, z) \rightarrow \D_{\FoG}^{2} (f, \sigma, \mu, k, z)  b(z)^2 \Bigl(1+\beta(z) \mu^2 \Bigr)^2 \P_{\delta\delta}(k,z) \label{PFoG} \, ,\ee
where the FoG factor can be chosen as either the Gaussian or the Lorentzian \cite{Peacock92, DavisPeebles, 0407214} 
\ba \D_{\FoG}^{\Gau} (f, \sigma,\mu, k, z) &=& \exp \Bigl[ -\fr{\sigma^{2}(z) f^{2}(z)k^{2} \mu^{2} }{2H^{2}(z)} \Bigr] \label{FoGGau} \, , \\
\text{D}_{\text{FoG}}^{\text{Lor}} (k,z, \sigma,\mu) &=& \Bigl(1 + \fr{\sigma^{2}(z) f^{2}(z)k^{2} \mu^{2} }{H^{2}(z)} \Bigr)^{-1/2} \label{FoGLor} \, , \\
\text{D}_{\text{FoG}}^{\text{Lorh}} (k,z, \sigma,\mu) &=& \Bigl( 1 + \fr{\sigma^{2}(z) f^{2}(z)k^{2} \mu^{2} }{2H^{2}(z)} \Bigr)^{-1/2}\label{FoGLorh} \, , \ea
where we also introduce the so-called ``dispersion model'' for the Lorentzian model given in \cite{0407214}. One can replace the term inside of the FoG effect facotrs as  
\be  \fr{x}{k} \equiv \fr{\sigma(z) f(z)}{H(z)} = \fr{\sigma_{0}}{D_{0} H_{0}} \fr{D(z) f(z)}{E(z)} = \fr{\sigma_{0} f_0}{H_{0}} \fr{D'(z)}{D'(z_0)} \fr{(1+z)}{E(z)} \label{FoGGau2} \, , \ee
where $\sigma(z) = \fr{D(z)}{D_{0}} \sigma_{0}$, $f(z) = \fr{a}{D(a)} \fr{d D}{da} = -\fr{(1+z)}{D(z)} D'(z)$, $D'(z) \equiv \fr{d D(z)}{dz}$, and $\E(z) \equiv \fr{H(z)}{H_{0}} = \sqrt{\Omega_{m0} (1+z)^3 + (1-\Omega_{m0}) (1+z)^{3(1+\omega_0)}}$. We assume the flat $\omega$CDM model with the constant dark energy equation of state, $\omega_{0}$ in $\E(z)$.

The linear theory multipoles with the Gaussian FoG factor of Eq. (\ref{FoGGau}) are given by
{\fontsize 80 \ba \P_{0}^{\lin} &=& - \frac{e^{-x^2} \left(6 f^2 x+4 f (2 b+f) x^3-e^{x^2} \sqrt{\pi } \left(3 f^2+4 b f x^2+4 b^2 x^4\right) \Erf[x]\right) \P_{\delta \delta }}{8 x^5} \label{P0lin} \, , \\
\P_{2}^{\lin} &=& -\frac{5 e^{-x^2} \Bigl(12 b^2 x^4+4 b f x^2 \left(9+4 x^2\right)+f^2 \left(45+24 x^2+8 x^4\right)\Bigr) \P_{\delta \delta }}{16 x^6} \nonumber \\ 
&+& \frac{5 \sqrt{\pi } \Bigl(45 f^2+6 (6 b-f) f x^2+4 b (3 b-2 f) x^4-8 b^2 x^6\Bigr) \Erf[x] P_{\delta \delta }}{32 x^7} \label{P2lin} \, , \\ 
\P_{4}^{\lin} &=& -\frac{9 e^{-x^2} \Bigl(20b^2 x^4\left(21+2 x^2\right)+4b f x^2\left(525+170 x^2+32 x^4\right)+f^2\left(3675+1550 x^2+416 x^4+64 x^6\right)\Bigr)\P_{\delta \delta }}{128 x^8} \nonumber \\ 
&+& \frac{27 \sqrt{\pi } \Bigl(4 b f x^2 \left(175-60 x^2+4 x^4\right)+4 b^2 x^4 \left(35-20 x^2+4 x^4\right)+f^2 \left(1225-300 x^2+12 x^4\right)\Bigr) \text{Erf}[x] \P_{\delta \delta }}{256 x^9} \label{P4lin} \, , \ea}
where $\Erf$ denotes the error function. Thus, the ratios of multipoles are given by
{\fontsize 80 \ba \R_{2}^{\lin} &\equiv& \fr{\P_{2}^{\lin}}{\P_{0}^{\lin}} = \frac{5 \Bigl(90 f^2 x+24 f (3 b+2 f) x^3+8 \left(3 b^2+4 b f+2 f^2\right) x^5\Bigr)}{4 x^2 \Bigl(6 f^2 x+4 f (2 b+f) x^3-e^{x^2} \sqrt{\pi } \left(3 f^2+4 b f x^2+4 b^2 x^4\right) \text{Erf}[x]\Bigr)} \nonumber \\
&-& \frac{5 e^{x^2} \sqrt{\pi } \Bigl(45 f^2+6 f (6 b-f) x^2+4 b (3 b-2 f) x^4-8 b^2 x^6\Bigr) \text{Erf}[x]}{4 x^2 \Bigl(6 f^2 x+4 f (2 b+f) x^3-e^{x^2} \sqrt{\pi } \left(3 f^2+4 b f x^2+4 b^2 x^4\right) \text{Erf}[x]\Bigr)} \label{R2lin} \, , \\
\R_{4}^{\lin} &\equiv& \fr{\P_{4}^{\lin}}{\P_{0}^{\lin}} = \frac{9 \Bigl(20 b^2 x^5 \left(21+2 x^2\right)+4 b f x^3 \left(525+170 x^2+32 x^4\right)+f^2 x \left(3675+1550 x^2+416 x^4+64 x^6\right)\Bigr)}{16 x^4 \Bigl(6 f^2 x+4 f (2 b+f) x^3-e^{x^2} \sqrt{\pi } \left(3 f^2+4 b f x^2+4 b^2 x^4\right) \text{Erf}[x]\Bigr)} \nonumber \\
&-& \frac{27 e^{x^2} \sqrt{\pi } \Bigl(4 b f x^2 \left(175-60 x^2+4 x^4\right)+4 b^2 x^4 \left(35-20 x^2+4 x^4\right)+f^2 \left(1225-300 x^2+12 x^4\right)\Bigr) \Erf[x]}{32 x^4 \Bigl(6 f^2 x+4 f (2 b+f) x^3-e^{x^2} \sqrt{\pi } \left(3 f^2+4 b f x^2+4 b^2 x^4\right) \Erf[x]\Bigr)} \label{R4lin}  \, . \ea}
All of these forms Eqs.(\ref{P0lin})-(\ref{R4lin}) are independent of the linear matter PS as in the Kaiser limit. Also, these results are equal to those of the Kaiser limit given by Eqs.(\ref{P0Kai})-(\ref{R4Kai}) when one adopts $x \rightarrow 0$ limit in the above Eqs.(\ref{P0lin})-(\ref{R4lin}). We show the multipoles and their ratios for the Lorentzian FoG factors of Eqs.(\ref{FoGLor}) and (\ref{FoGLorh}) in the Appendix. As in the Kaiser limit case, one can obtain the real-space linear PS, $\P_{\delta\delta}$ from the linear combinations of the multipoles  
{\fontsize 80 \ba \P_{\delta\delta} &=& \P_{0}^{\lin} + c_{2}^{\lin} \P_{2}^{\lin} + c_{4}^{\lin} \P_{4}^{\lin} \label{Pddder} \, , \\
c_{4}^{\lin} &=&  \fr{-\left(16 x^3 \left(4 x^2 \left(4 b f x^2+4 e^{x^2} x^4+f^2 \left(3+2 x^2\right)\right)+5 c_{2}^{\lin} \left(12 b^2 x^4+4 b f x^2 \left(9+4 x^2\right)+f^2 \left(45+24 x^2+8 x^4\right)\right)\right)\right)}{c_{2}^{\lin}} \label{c4} \\ 
&-&  \fr{ \left(8 e^{x^2} \sqrt{\pi } x^2 \left(-4 \left(3 f^2 x^2+4 b f x^4+4 b^2 x^6\right)+5 c_{2}^{\lin} \left(4 b f x^2 \left(-9+2 x^2\right)+4 b^2 x^4 \left(-3+2 x^2\right)+f^2 \left(-45+6 x^2\right)\right)\right) \text{Erf}[x]\right)}{c_{2}^{\lin}}  \nonumber \, , \\ 
c_{2}^{\lin} &=& 18 x \left(20 b^2 x^4 \left(21+2 x^2\right)+4 b f x^2 \left(525+170 x^2+32 x^4\right)+f^2 \left(3675+1550 x^2+416 x^4+64 x^6\right)\right) \nonumber \\ 
&-& 27 e^{x^2} \sqrt{\pi } \left(4 b f x^2 \left(175-60 x^2+4 x^4\right)+4 b^2 x^4 \left(35-20 x^2+4 x^4\right)+f^2 \left(1225-300 x^2+12 x^4\right)\right) \text{Erf}[x] \label{c2} \, . \ea}
The above Eq.(\ref{Pddder}) is the extended version of Eq.(\ref{PddKai}) when one includes the FoG effect in the multipoles.

One can extend the linear matter PS to the quasi-linear one by using the standard perturbation theory. If one uses the approximation of the redshift-space PS up to only $l = 0, 2, 4$ multipoles, then one can always write \cite{0407214, 14077325}
\be \P_{s}(k) = \P(k) \Bigl[ 1 + 2 A_{2}(k) \mu^2 + A_{4}(k) \mu^4 \Bigr] \label{PSco} \, , \ee
where $A_{2}$ and $A_{4}$ are obtained from the perturbation theories. This approximation in Eq.(\ref{PSco}) is also valid in the quasi-linear regime because the multipoles higher than $l=4$ are generated only for $k \geq 0.2 \Mpch$ \cite{0407214}.

\section{Quasi-Linear Scale}
\label{QLS}

DR12 uses the TNS model for the anisotropic galaxy power spectrum in the analysis \cite{TNS} 
\ba \P_{g}^{(\TNS)}(k,\mu) &=& \exp{-(f k \mu \sigma_{v})^2} \Bigl[\P_{g,\delta\delta}(k) + 2 f \mu^2 \P_{g,\delta\theta}(k) + f^2 \mu^4 \P_{\theta\theta}(k) + b_{1}^3A(k,\mu,\beta)+b_1^4 B(k,\mu,\beta) \Bigr] \nonumber \\
&\equiv& \exp{-(f k \mu \sigma_{v})^2} \Bigl[\P_{g,\delta\delta}(k) + \mu^2 {\cal A} (k) +\mu^4 {\cal B}(k) + \mu^{6} {\cal C}(k) + \mu^{8} {\cal D}(k) \Bigr] \label{PgTNS} \\
&\simeq& \exp{-(f k \mu \sigma_{v})^2} \Bigl[\P_{g,\delta\delta}(k) + \mu^2 {\cal A} (k) +\mu^4 {\cal B}(k) \Bigr] \label{PgTNSapp}  \ea
where
\ba A(k,\mu,b) &=& (k \mu f) \int \fr{d^3q}{(2\pi)^3} \fr{q_{z}}{q^2} \Bigl[ B_{\sigma}(\vec{q}, \vec{k}-\vec{q}, -\vec{k}) - B_{\sigma}(\vec{q}, \vec{k}, -\vec{k}-\vec{q}) \Bigr] \label{A} \, , \\
B(k,\mu,b) &=& (k \mu f)^2  \int \fr{d^3q}{(2\pi)^3}  F(\vec{q}) F(\vec{k}-\vec{q}) \label{B} \, , \\
(2\pi)^{3} \delta_{\D}(\vec{k}_{123}) B_{\sigma}(\vec{k}_1,\vec{k}_2,\vec{k}_3) &\equiv& \Biggl< \theta(\vec{k}_1) \Bigl[ b(k_2) \delta(\vec{k}_2) + f \fr{k_{2z}^{2}}{k_{2}^2} \theta(\vec{k}_{2}) \Bigr] \Bigl[ b(k_3) \delta(\vec{k}_3) + f \fr{k_{3z}^{2}}{k_{3}^2} \theta(\vec{k}_{3}) \Bigr] \Biggr> \label{Bsigma} \, , \\
F(\vec{q}) &\equiv& \fr{q_{z}}{q^2} \Bigl[ b(q) \P_{\delta\theta}(q) + f \fr{q_{z}^2}{q^2} \P_{\theta\theta} (q) \Bigr] \label{F}  \, . \ea
In order to obtain the ratios of multipoles up to hexadecapole, one needs to consider the terms up to $\mu^4$ in the above Eq.(\ref{PgTNS}) and thus we use the approximation of it given by Eq.(\ref{PgTNSapp}). This approximation is also used in the data analysis \cite{160703150}. It has been shown that the recovering the real-space PS from the redshift-space multipoles is still given by the same linear combination as in the Kaiser limit up to $k < 0.2 \, h$Mpc$^{-1}$ and this fact rationalizes this approximation \cite{0111575, 0310725, 0407214}.   
  
From the Eq.(\ref{PgTNSapp}), one obtains the multipoles
 {\fontsize 80  \ba  \P_{0}^{\Quas} &=& \frac{-e^{-x^2} \Bigl(3\B +2 (\A+\B) x^2\Bigr)}{4x^4} + \frac{\sqrt{\pi } \text{Erf}[x] \left(3 \B+2 \A x^2+4 x^4 \P_{g,\delta \delta }\right)}{8 x^5} \label{P0q} \, , \\ 
 \P_{2}^{\Quas} &=& \frac{-5 e^{-x^2} \left(45 {\cal B}+6 (3 {\cal A}+4 {\cal B}) x^2+8 ({\cal A}+{\cal B}) x^4+12 x^4 \P_{g, \delta \delta }\right)}{16 x^6}\nonumber \\
 &+& \frac{5 \sqrt{\pi } \text{Erf}[x] \Bigl(45 {\cal B}+6 (3 {\cal A}-{\cal B}) x^2-4 {\cal A} x^4+4 x^4 \left(3-2 x^2\right) \P_{g, \delta \delta }\Bigr)}{32 x^7} \label{P2q} \, , \\
 \P_{4}^{\Quas} &=& \frac{-9 e^{-x^2} \left(3675{\cal B}+50 (21{\cal A}+31{\cal B})x^2+4 (85{\cal A}+104{\cal B})x^4+64 ({\cal A}+{\cal B}) x^6+20 x^4 \left(21+2 x^2\right) \P_{g, \delta \delta }\right)}{128 x^8} \nonumber \\ 
 &+& \frac{27\sqrt{\pi } \text{Erf}[x] \left(1225 \B+50 (7 \A-6 \B) x^2-12 (10 \A-\B) x^4+8 \A x^6+4 x^4 \left(35-4 x^2 \left(5-x^2\right)\right) \P_{g, \delta \delta }\right)}{256 x^9} \label{P4q} \, .  \ea}
 From Eqs.(\ref{P0q})-(\ref{P4q}), the ratios of the multipoles are given by
 {\fontsize 80 \ba \R_{2}^{\Quas} &=& \frac{5 \Bigl(45 \B+6 (3 \A+4 \B) x^2+8 (\A+\B) x^4+12 x^4 \P_{g,\delta \delta }\Bigr)}{2 x \Bigl(6 \B x+4 (\A+\B) x^3-e^{x^2} \sqrt{\pi } \text{Erf}[x] \left(3 \B+2 \A x^2+4 x^4 \P_{g,\delta \delta }\right)\Bigr)} \nonumber \\ 
&+& \frac{5 e^{x^2} \sqrt{\pi } \text{Erf}[x] \Bigl(-45 \B-6 (3 \A-\B) x^2+4 \A x^4-4 x^4 \left(3-2 x^2\right) \P_{g,\delta \delta }\Bigr)}{4 x^2 \Bigl(6 \B x+4 (\A+\B) x^3-e^{x^2} \sqrt{\pi } \text{Erf}[x] \left(3 \B+2 \A x^2+4 x^4 \P_{g,\delta \delta }\right)\Bigr)}  \label{R2q} \, , \\
 \R_{4}^{\Quas} &=& \frac{9 \Bigl(3675 \B+50 (21 \A+31 \B) x^2+4 (85 \A+104 \B) x^4+64 (\A+\B) x^6+20 x^4 \left(21+2 x^2\right) \P_{g,\delta \delta }\Bigr)}{16 x^3 \Bigl(6 \B x+4 (\A+\B) x^3-e^{x^2} \sqrt{\pi } \text{Erf}[x] \left(3 \B+2 \A x^2+4 x^4 \P_{g,\delta \delta }\right)\Bigr)} \nonumber \\ 
&-& \frac{27 e^{x^2} \sqrt{\pi } \text{Erf}[x] \Bigl(1225 \B+50 (7 \A-6 \B) x^2-12 (10 \A-\B) x^4+8 \A x^6+4 x^4 \left(35-4 x^2 \left(5-x^2\right)\right) \P_{g,\delta \delta }\Bigr)}{32 x^4 \Bigl(6 \B x+4 (\A+\B) x^3-e^{x^2} \sqrt{\pi } \text{Erf}[x] \left(3 \B+2 \A x^2+4 x^4 \P_{g,\delta \delta }\right)\Bigr)} \label{R4q} \, .
\ea}
One can extract $\P_{g,\delta\delta}$, $\A$, and $\B$ from observed multipoles values  by using the Eqs.(\ref{P0q})-(\ref{P4q}),
 \ba \P_{g,\delta\delta} &=& \fr{ \N_{\P_{\delta\delta}\P_{0}} +\N_{\P_{\delta\delta}\P_{2}} +\N_{\P_{\delta\delta}\P_{4}} }{\D_{\P} } \label{Pdeltadelta} \, , \\
{\cal A} &=& \fr{ \N_{{\cal A} \P_{0}} +\N_{{\cal A} \P_{2}} +\N_{{\cal A} \P_{4}} }{\D_{\P} } \label{calA} \, , \\
{\cal B} &=& \fr{ \N_{{\cal B} \P_{0}} +\N_{{\cal B} \P_{2}} +\N_{{\cal B}\P_{4}} }{\D_{\P} } \label{calB} \, ,  \ea
where
{\fontsize 80 \ba \N_{\P_{\delta\delta}\P_{0}} &=&  21 e^{x^2} x \left(4 x^2 \left(675+600 x^2+16 x^4+8 x^6\right)+4 e^{x^2} \sqrt{\pi } x \left(-675-150 x^2+124 x^4-44 x^6+8 x^8\right) \text{Erf}[x] \right)  \P_{0} \nonumber \\ &+& 21 e^{x^2} x \left( 3 e^{2 x^2} \pi  \left(225-100 x^2+12 x^4\right) \text{Erf}[x]^2\right) \P_0 \label{NPdeltadeltaP0} \, , \\
\N_{\P_{\delta\delta}\P_{2}}  &=&  24 e^{x^2} x^3 \left(2 x^2 \left(-210+x^2+2 x^4\right)+e^{x^2} \sqrt{\pi } x \left(420-37 x^2+68 x^4+4 x^6\right) \text{Erf}[x] \right) \P_{2} \nonumber  \\
&+&  24 e^{x^2} x^3 \left( 3 e^{2 x^2} \pi  \left(-35+6 x^2\right) \text{Erf}[x]^2\right) \P_2 \label{NPdeltadeltaP2} \, ,\\
\N_{\P_{\delta\delta}\P_{4}}  &=&  32 e^{x^2} x^5 \left(4 x^2 \left(3-x^2\right)-2 e^{x^2} \sqrt{\pi } x \left(6-x^2+2 x^4\right) \text{Erf}[x]+3 e^{2 x^2} \pi  \text{Erf}[x]^2\right) \P_4 \label{NPdeltadeltaP4}  \, , \\
\N_{\A \P_0} &=& -84 e^{x^2} x^3 \left(4 x^2 \left(225+60 x^2+6 x^4+4 x^6\right)+4 e^{x^2} \sqrt{\pi } x \left(-225+90 x^2+29 x^4+8 x^6+4 x^8\right) \text{Erf}[x] \right) \P_{0} \nonumber \\ 
&-& 84 e^{x^2} x^3 \left(3 e^{2 x^2} \pi  \left(75-80 x^2+12 x^4\right) \text{Erf}[x]^2\right) \P_0  \label{NAP0} \, , \\
\N_{\A \P_2} &=& -24 e^{x^2} x^5 \left(2 x^2 \left(63+48 x^2+4 x^4\right)+e^{x^2} \sqrt{\pi } x \left(609+298 x^2+100 x^4+8 x^6\right) \text{Erf}[x] \right) \P_2 \nonumber  \\ 
&+& 24 e^{x^2} x^5 \left( 24 e^{2 x^2} \pi  \left(14- 3 x^2\right) \text{Erf}[x]^2\right) \P_2 \label{NAP2} \, , \\
\N_{\A \P_4} &=& 64 e^{x^2} x^7 \left(6 x^2+4 x^4+e^{x^2} \sqrt{\pi } x \left(9+8 x^2+4 x^4\right) \text{Erf}[x]-6 e^{2 x^2} \pi  \text{Erf}[x]^2\right) \P_4 \label{NAP4} \, , \\
\N_{\B \P_0} &=& -84 e^{x^2} x^5 \left(4 x^2 \left(45+4 x^4\right)+4 e^{x^2} \sqrt{\pi } x \left(-45+30 x^2+2 x^4+4 x^6\right) \text{Erf}[x] \right) \P_0 \nonumber \\ 
&-& 84 e^{x^2} x^5 \left(3 e^{2 x^2} \pi  \left(15-20 x^2+4 x^4\right) \text{Erf}[x]^2\right) \P_0 \label{NBP0} \, , \\
\N_{\B \P_2} &=& -48 e^{x^2} x^7 \left(42 x^2+4 x^4+e^{x^2} \sqrt{\pi } x \left(63+44 x^2+4 x^4\right) \text{Erf}[x]-6 e^{2 x^2} \pi  \left(7-2 x^2\right) \text{Erf}[x]^2\right) \P_2 \label{NBP2} \, , \\
\N_{\B \P_4} &=& 128 e^{x^2} x^9 \left(2 x^2+e^{x^2} \sqrt{\pi } x \left(1+2 x^2\right) \text{Erf}[x]-e^{2 x^2} \pi  \text{Erf}[x]^2\right) \P_4 \label{NBP4} \, , \\
\D_{\P} &=& 315 \left(4 x^3 (21+2 x^2)-4 e^{x^2} \sqrt{\pi } x^2 \left(9-28 x^2\right) \text{Erf}[x]-e^{2 x^2} \pi  x \left(27+58 x^2-20 x^4+8 x^6\right) \text{Erf}[x]^2 \right) \,  \nonumber  \\
&+& 315 \left( 12 e^{3 x^2} \pi ^{3/2} \text{Erf}[x]^3\right) \, . \ea}
One can obtain the ratios of multipoles of Eqs.(\ref{R2q}) and (\ref{R4q}) by using the measurements of multipoles in the reference \cite{160703150}. We compare these ratios of multipoles with those of liner theory. We constrain our analysis only for the North Galactic Cap (NGC) with including hexadecapole data. There are nine $k$-values in DR12 for the hexadecapole.  

\begin{figure}
\centering
\vspace{1.5cm}
\epsfig{file=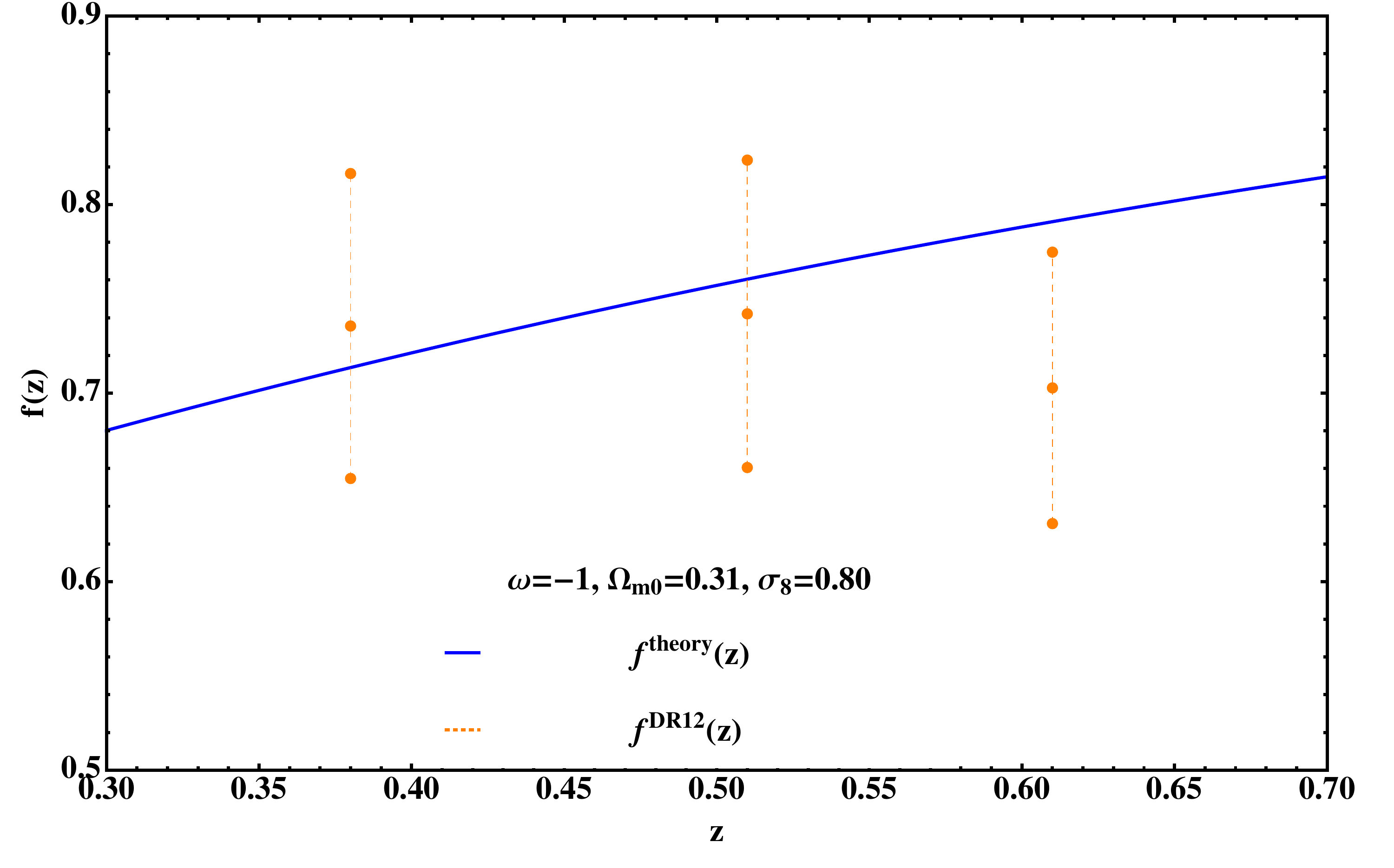,width=0.8\linewidth,clip=} 
\vspace{-0.5cm}
\caption{The evolution of the growth index, $f(z)$. The solid line depicts the theoretical values of $f(z)$ based on the fiducial model, $\Lambda$CDM with $\Omega_{\m 0} =0.31$ and $\sigma_{8}(z_0)=0.80$. The dashed lines represent the DR12 data at three different redshifts.} \label{fig1}
\end{figure}

First, we show the consistency of BOSS DR12 results for the growth index, $f$ for three redshift bins. One can use the theoretical expectation values of $\sigma_{8}(z) = (D_{1}(z)/D_{1}(z_0)) \sigma_{8}(z_0)$ to obtain $f(z)$ from the observed $f(z) \sigma_{8}(z)$ values. We show this in Fig.\ref{fig1}. The solid line represents the theoretical values of $f(z)$ based on the fiducial model, $\Lambda$CDM with $\Omega_{\m0} = 0.31$ and $\sigma_{8}(z_0) = 0.8$. The vertical dashed lines depict the DR12 data at three effective redshifts, $z = 0.38, 0.51$, and 0.61. $f(z)$ value at $z_{\eff} = 0.61$ deviates from the theoretical one by about 1.4-$\sigma$ as indicated in the reference \cite{160703150}. We adopt the fiducial cosmological parameters, which is used in the reference \cite{160703150} to generate the linear power spectrum model,  a flat $\Lambda$CDM model with $\Omega_{\m0} = 0.31, \Omega_{b}h^2 = 0.022, h = 0.676, \sigma_{8} = 0.8, n_{s} = 0.96, \sum m_{\nu} = 0.06$ eV. 

Now we investigate the ratios of multipoles at three effective redshifts by using the measurement of multipoles at the different k-values. 

\subsection{At $z_{1} = 0.38$}

\begin{figure}
\centering
\vspace{1.5cm}
\begin{tabular}{cc}
\epsfig{file=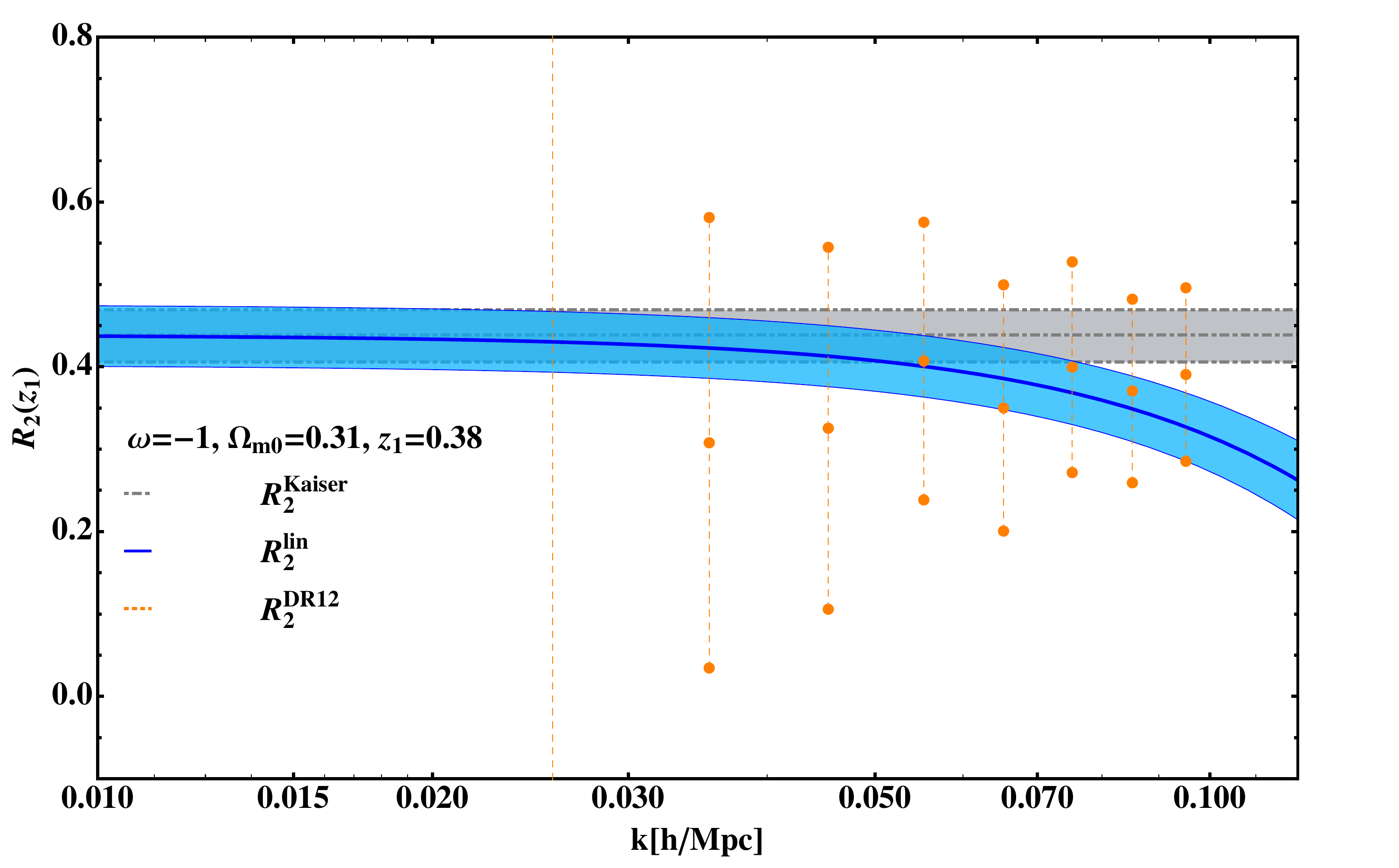,width=0.5\linewidth,clip=} &
\epsfig{file=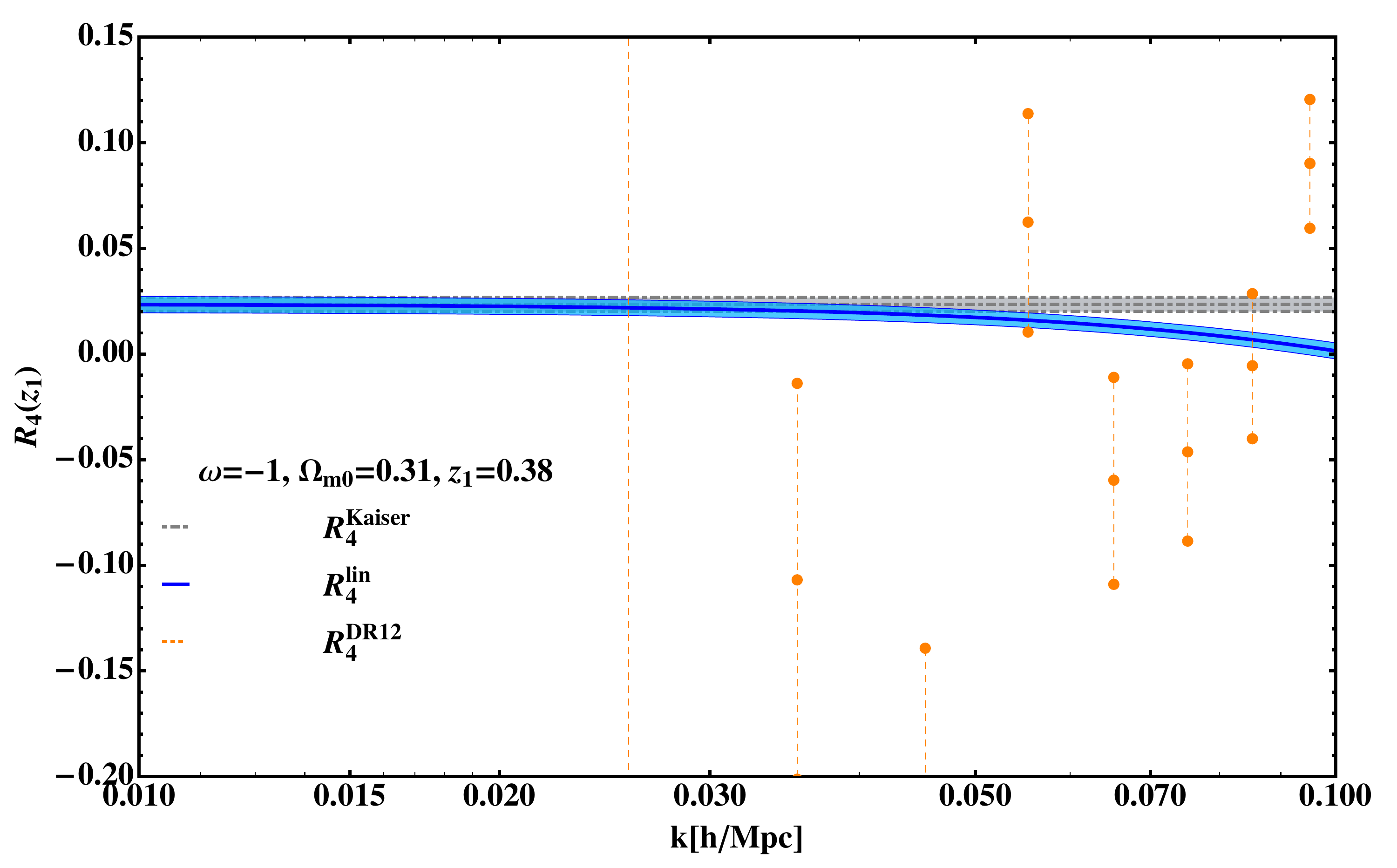,width=0.5\linewidth,clip=} \\
\end{tabular}
\vspace{-0.5cm}
\caption{The values of $\R_{2}$ and $\R_{4}$ at $z = 0.38$. a) The ratio of quadrupole to monopole, $\R_{2}$. The dark shaded region is the 1-$\sigma$ prediction of the Kaiser limit. The bright shaded region is the 1-$\sigma$ prediction of the linear theory. The vertical dashed lines indicate the 1-$\sigma$ results of the DR12.  b) The ratio of hexadecapole to monopole, $\R_{4}$ with the same notation as in the left panel. } \label{fig2}
\end{figure}

For the effective redshift $z_{1} = 0.38$, one can obtain the ratio of the quadrupole (hexadecapole) to the monopole moment of the redshift-space PS, $\R_{2} (\R_{4})$ from the DR12 results by using the analytic formulae given in Eqs.(\ref{R2q}) and (\ref{R4q}). The left panel of Fig.\ref{fig2} depicts both the linear $R_{2}$ and the quasi-linear one. The dark shaded region indicates the 1-$\sigma$ theoretical prediction for the ratio of the quadrupole to the monopole in the Kaiser limit, $\R_{2}^{\Kai}$ given by Eq.(\ref{R2Kai}). BOSS DR12 results provide both $f(z_{1}) \sigma_{8}(z_{1})$ and $b(z_1) \sigma_{8}(z_{1})$ with the present fiducial value of $\sigma_{8}(z_0) = 0.8$. We derive both $f(z_1)$ and $b(z_1)$ with 1-$\sigma$ error to obtain the corresponding $\R_{2}^{\Kai}(z_1)$. $R_{2}^{\Kai} = 0.438 \pm 0.032$ for any given value of $k$. The bright shaded region shows the 1-$\sigma$ region of the linear theory prediction for the ratio of the quadrupole to the monopole, $\R_{2}^{\lin}$ including the FoG effects. By including the FoG effects, $R_{2}$ depends on $k$ as given in Eq.(\ref{R2lin}). For $k < 0.03 \Mpch$ (large scales), $\R_{2}^{\lin}$ is almost same as $\R_{2}^{\Kai}$ as expected. The small scale damping effect becomes important from $k > 0.03 \Mpch$. $\R_{2}^{\lin} = 0.327 \pm 0.042$ at $k = 0.095 \Mpch$. The 1-$\sigma$ values of $\R_{2}$ obtained from the DR12 multipoles are depicted by the vertical dashed lines in the left panel of Fig.\ref{fig2}. All of the measured $\R_{2}$ values are consistent with the linear theory predictions except the one at the largest scale, $k = 0.016$ $\Mpch$. Both the covariance matrix and the $\chi^2$-value of $R_{2}^{\DR12}(z_1)$ are shown in the appendix \ref{Appendix}. The covariance matrix of $\R_{2}^{\DR12}$ is given by Eq.(\ref{CR2z1}). The degrees of freedom (d.o.f) at this redshift are nine. $\chi^2$-value is 6.10 shown in the table.\ref{tab-1}. We also show the linear theory predictions and measured values for the ratio of the hexadecapole to monopole, $\R_{4}$ in the right panel of Fig.\ref{fig2}. The dark shaded region is the 1-$\sigma$ Kaiser limit prediction, $\R_{4}^{\Kai}$ given by Eq.(\ref{R4Kai}). These values range from 0.021 to 0.027 for the measured values of $f$ and $b$ at $z=0.38$. The bright shaded region is the 1-$\sigma$ region of the linear theory prediction with the FoG effect, $\R_{4}^{\lin}$. The measured values of $\R_{4}^{\text{DR}12}$ are sparse and the most of them are deviated from the prediction of the linear theory as shown in the right panel of Fig.\ref{fig2}. The covariance matrix of $R_{4}^{\DR12}$ at this redshift is given in Eq.(\ref{CR4z1}). The $\chi^2$ value is 131 as shown in the table.\ref{tab-1}.

\subsection{At $z_{2} = 0.51$}
\begin{figure}
\centering
\vspace{1.5cm}
\begin{tabular}{cc}
\epsfig{file=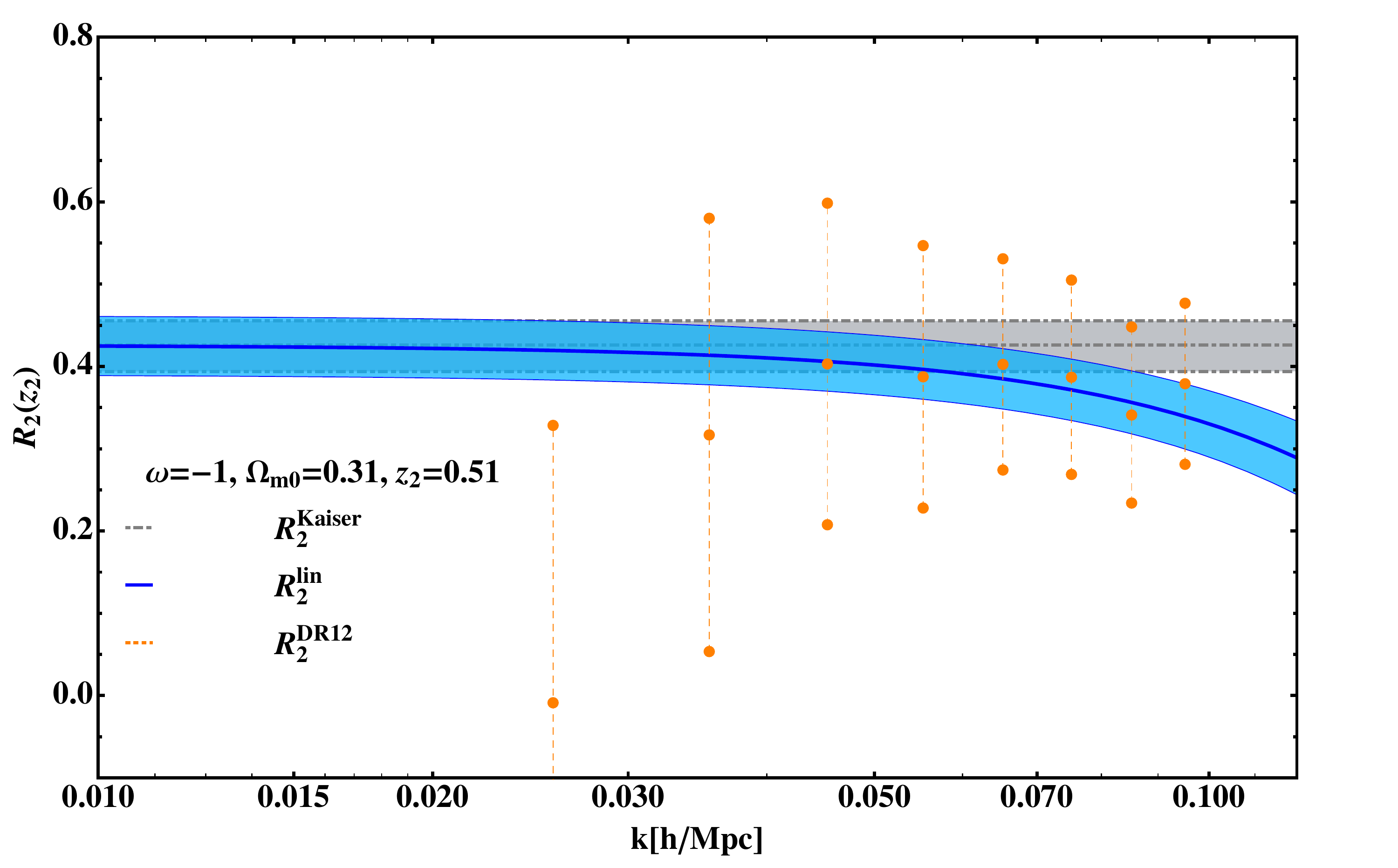,width=0.5\linewidth,clip=} &
\epsfig{file=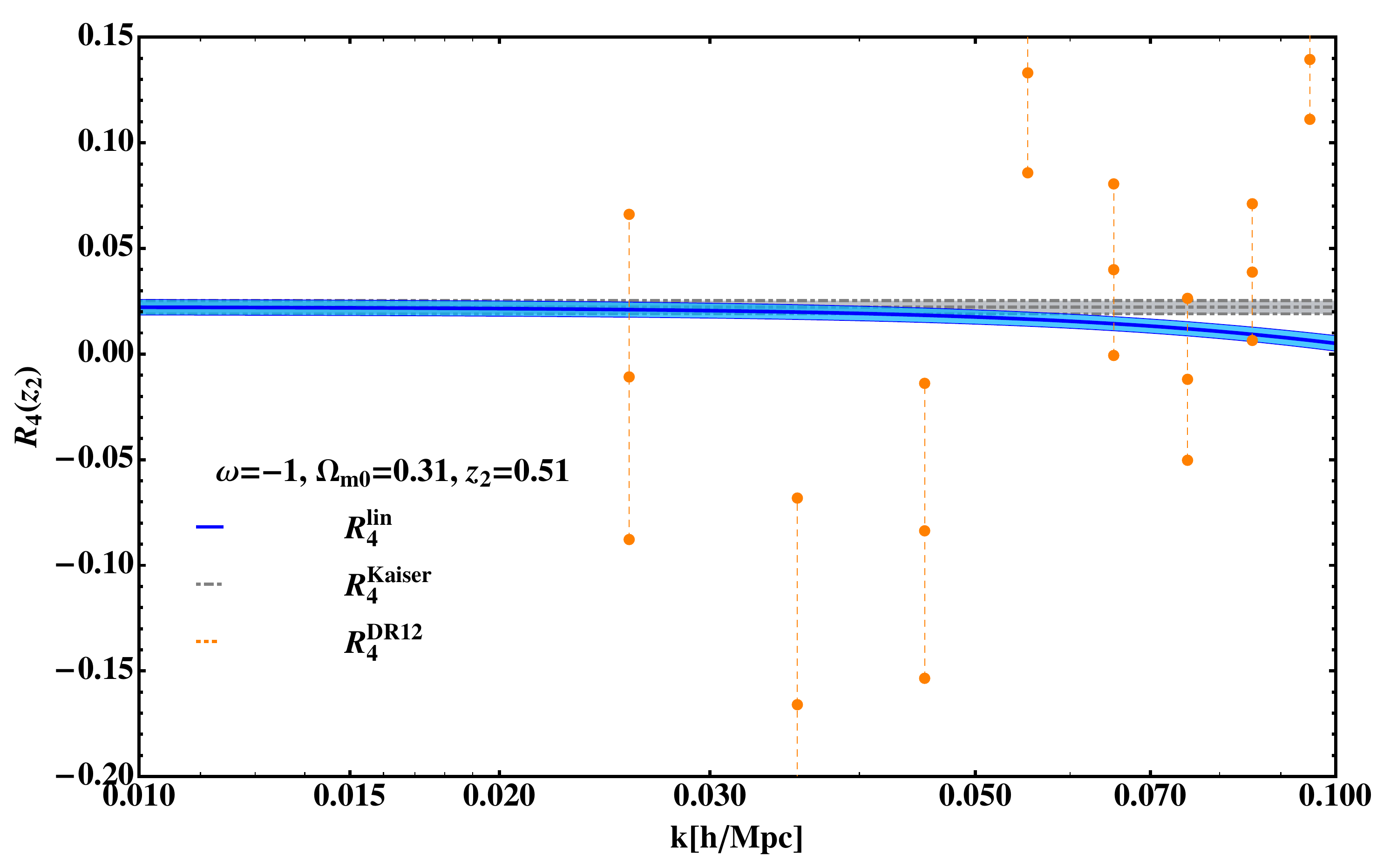,width=0.5\linewidth,clip=} \\
\end{tabular}
\vspace{-0.5cm}
\caption{The values of $\R_{2}$ and $\R_{4}$ at $z = 0.51$. a) $\R_{2}$ with the same notations as those of Fig.\ref{fig2}. b) $\R_{4}$ with the same notations as those of Fig.\ref{fig2}.} \label{fig3}
\end{figure}

We repeat the same analyses for $\R_{2}$ and $\R_{4}$ at the effective redshift $z_{2} = 0.51$. The left panel of Fig.\ref{fig3} depicts the $\R_{2}$-values of the different models and those of the measured one. The Kaiser limit prediction of $\R_{2}$ is $0.426 \pm 0.032$ represented by the dark shaded region. The bright shaded region indicates the 1-$\sigma$ prediction for the linear theory with the FoG effect, $\R_{2}^{\lin}$. Again both $\R_{2}^{\lin}$ and $\R_{2}^{\Kai}$ are consistent with each other up to $k < 0.04 \Mpch$. At $k= 0.095 \Mpch$, the values of $\R_{2}^{\lin}$ range from 0.300 to 0.379 within 1-$\sigma$. The vertical dashed lines are 1-$\sigma$ values of $\R_{2}$ obtained from DR12 multipoles. All of the measured $\R_{2}$-values are consistent with the linear theory prediction except for the largest scales at $k = 0.016 \Mpch$. The measured $R_{2}$ value at $k=0.016$ is undetermined and thus the d.o.f at this redshifts are eight instead of nine. The covariance matrix of $R_{2}^{\DR12}$ at $z_{2}$ is given in Eq.(\ref{CR2z2}). $\chi^2$-value of $R_{2}(z_2)$ is 2.16. We also show the $\R_{4}$ in the right panel of Fig.\ref{fig3}. The dark shaded region is $\R_{4}^{\Kai}$. $\R_{4}^{\Kai} = 0.022 \pm 0.003$. The bright shaded region is the 1-$\sigma$ prediction for $\R_{4}^{\lin}$ ranged from 0.003 to 0.010 at $k = 0.095 \Mpch$. The measured values of $\R_{4}^{\DR12}$ are sparse as shown in the right panel of Fig.\ref{fig3}. The $\chi^2$-value is 34.78. 

\subsection{At $z_{3} = 0.61$}
\begin{figure}
\centering
\vspace{1.5cm}
\begin{tabular}{cc}
\epsfig{file=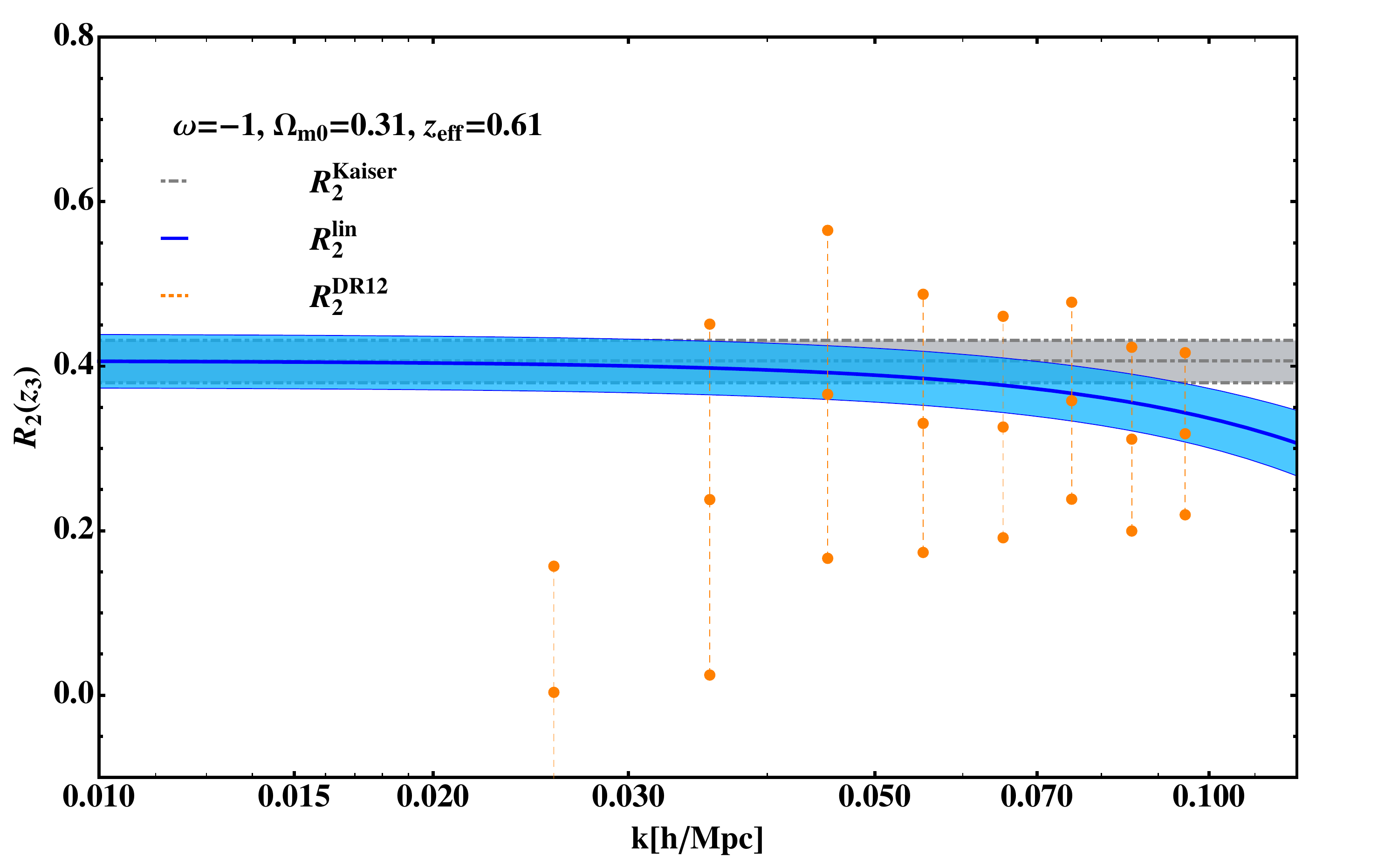,width=0.5\linewidth,clip=} &
\epsfig{file=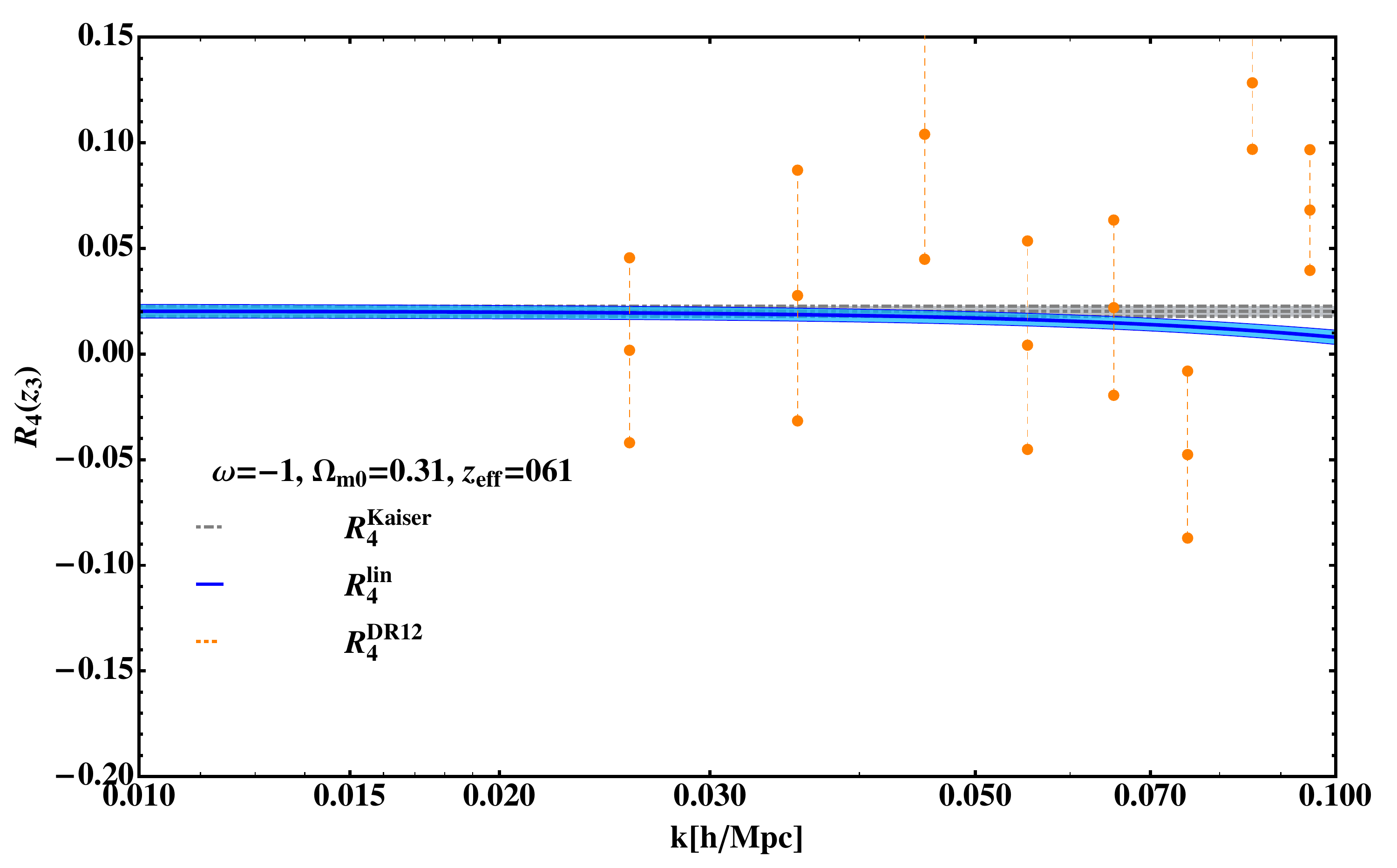,width=0.5\linewidth,clip=} \\
\end{tabular}
\vspace{-0.5cm}
\caption{The values of $\R_{2}$ and $\R_{4}$ at $z = 0.61$. a) $\R_{2}$ with the same notations as those of Fig.\ref{fig2}. b) $\R_{4}$ with the same notations as those of Fig.\ref{fig2}. } \label{fig4}
\end{figure}

The theory predictions and the measured values of both $\R_{2}$ and $\R_{4}$ at the effective redshift $z_{3} = 0.61$ are shown in Fig.\ref{fig4}. The left panel of Fig.\ref{fig4} shows $\R_{2}^{\Kai}$, $\R_{2}^{\lin}$, and $\R_{2}^{\DR12}$ at this redshift. $R_{2}^{\DR12}$ The results are similar to those at $z_{2}$ with slightly worse consistent between theory and observation. The covariance matrix for $\R_{2}^{\DR12}$ is given in Eq.(\ref{CR2z3}). $\chi^2 = 7.90$. Even though the measured $\R_{4}^{\DR12}$ shows the better consistency with the linear theory prediction than those at the other redshift, this might be due to the measurement error of the growth index, $f(z)$ at this redshift. This fact is shown in the right panel of Fig.\ref{fig4}. We also provide the covariance matrix for $R_{4}$ in Eq.(\ref{CR4z3}). $\chi^2$ value is 22.38.   

\section{Recovery of the Real-Space Matter Power Spectrum}
\label{RSMPS} 
\begin{figure}
\centering
\vspace{1.5cm}
\begin{tabular}{ccc}
\epsfig{file=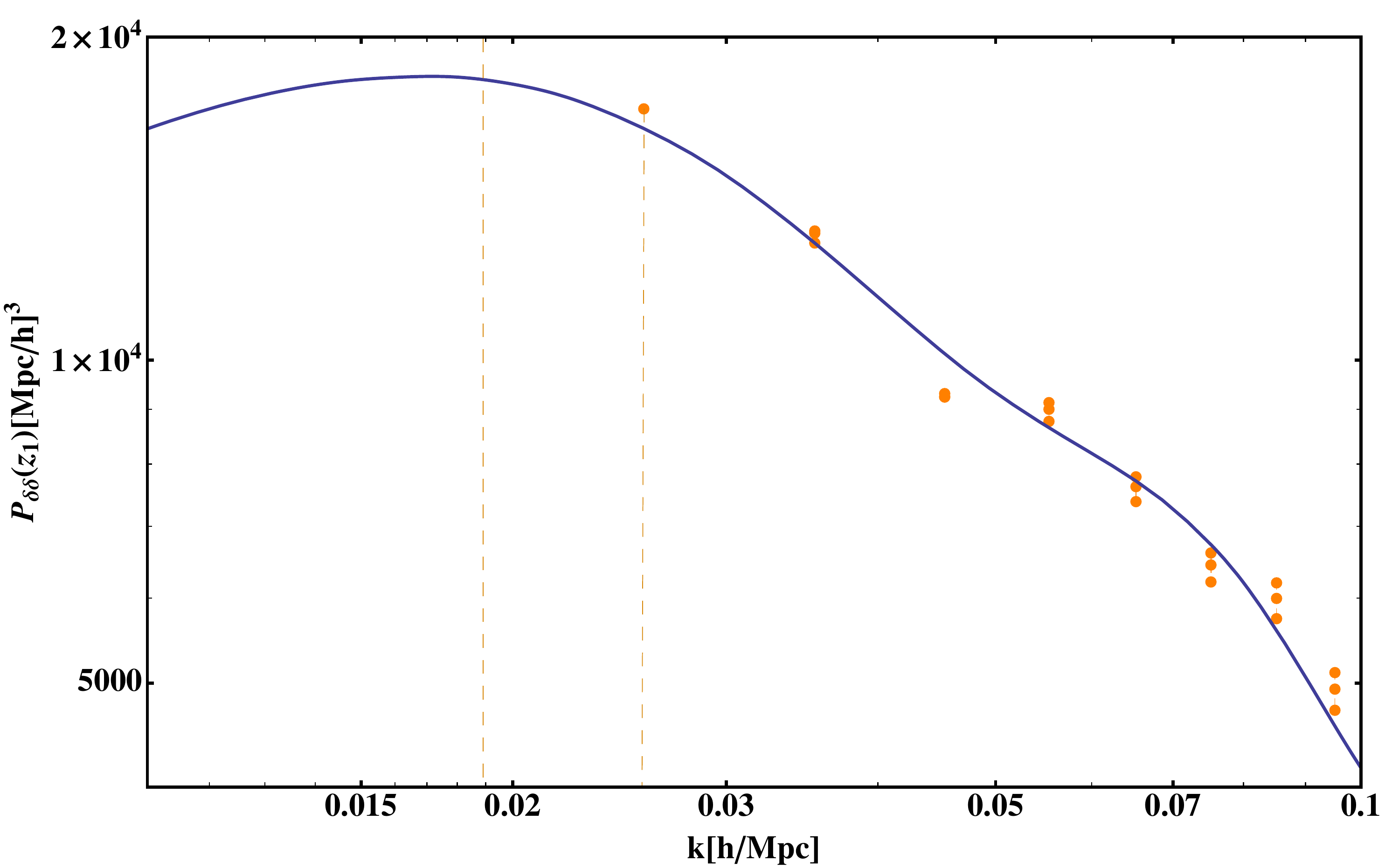,width=0.33\linewidth,clip=} &
\epsfig{file=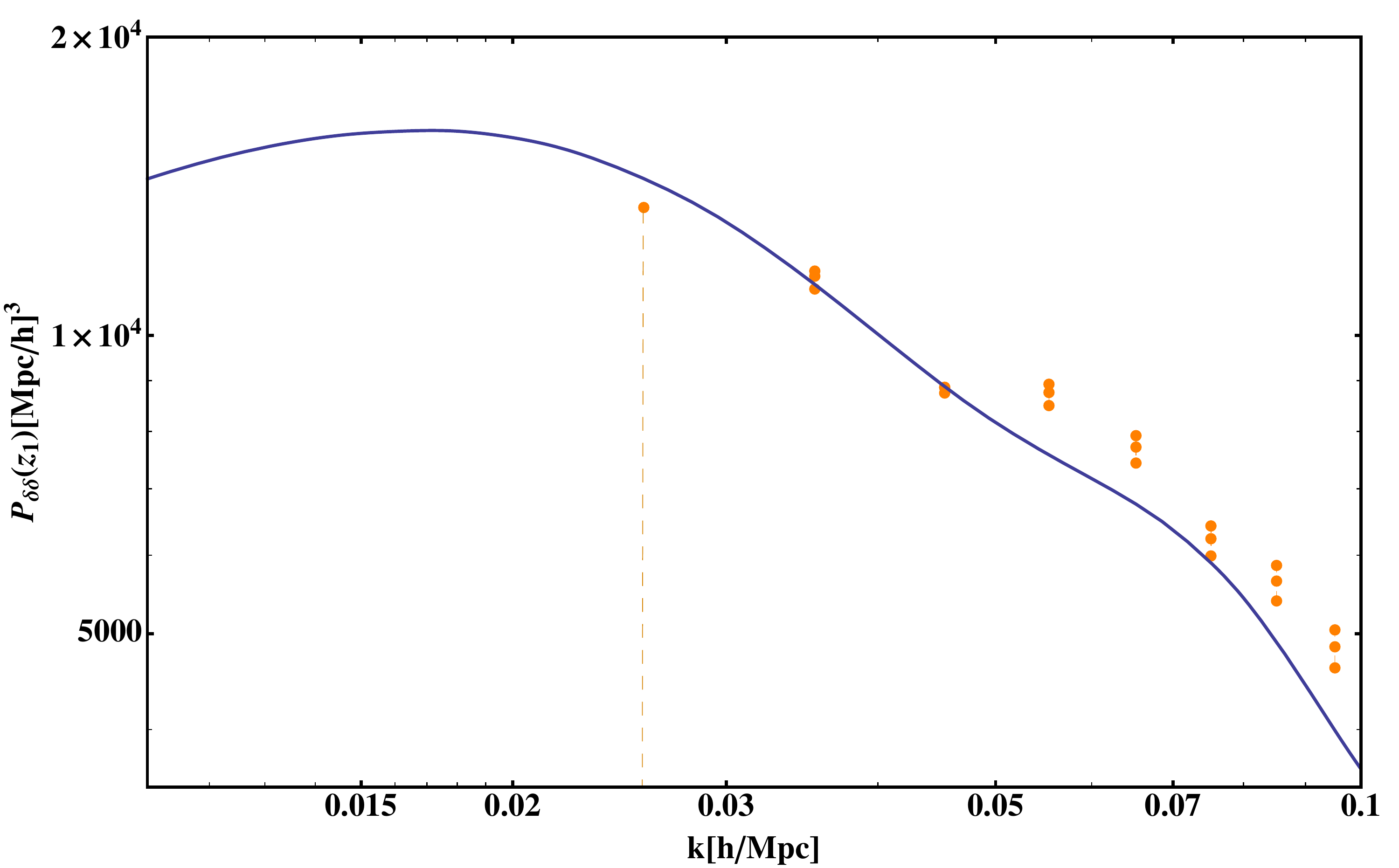,width=0.33\linewidth,clip=} & 
\epsfig{file=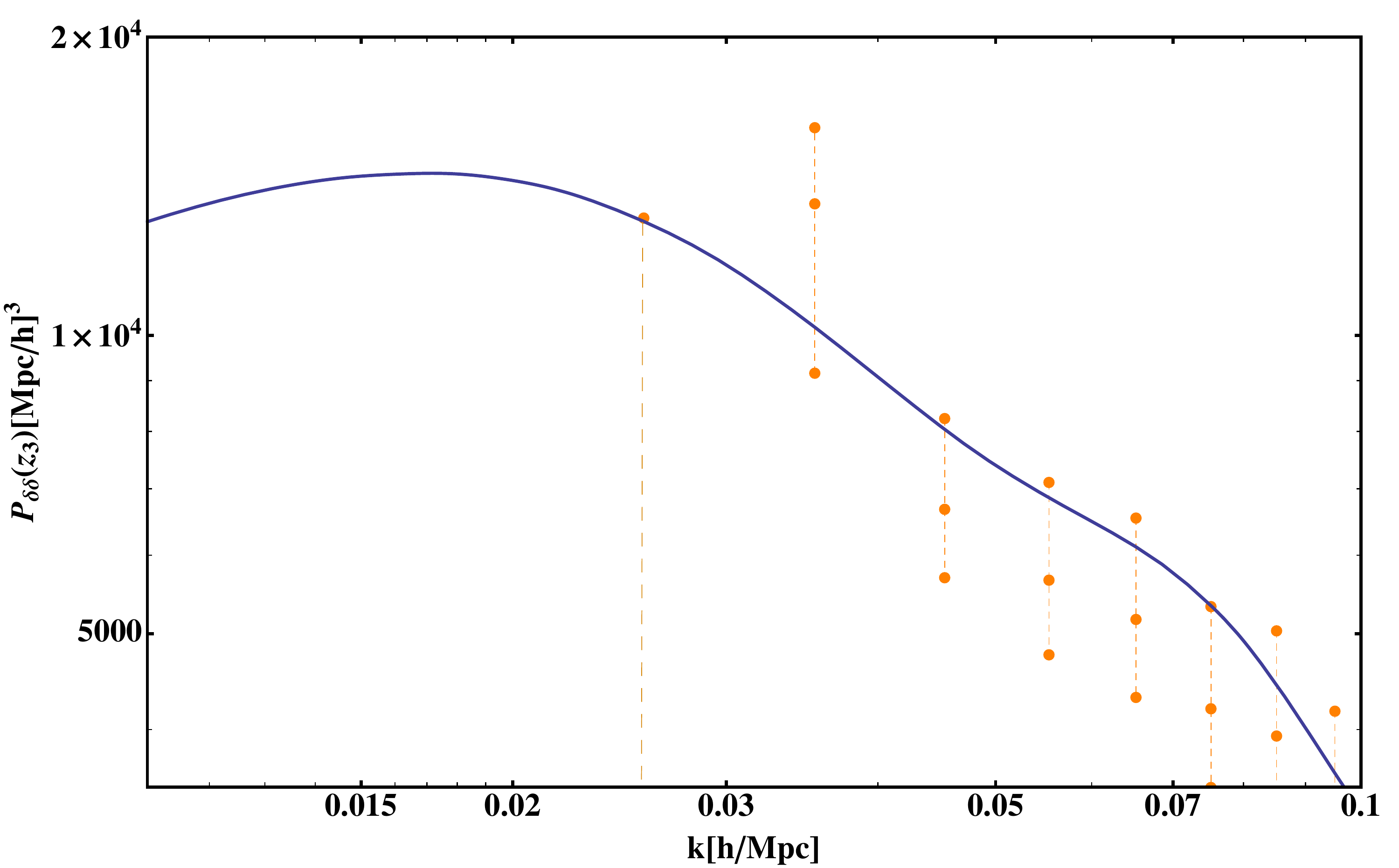,width=0.33\linewidth,clip=}\\
\end{tabular}
\vspace{-0.5cm}
\caption{The recovered matter power spectra at $z _1, z_2$, and $z_3$. a) The solid line is the linear matter power spectrum of the fiducial model at $z_1$. The vertical lines are 1-$\sigma$ measurements of the recovered power spectrum. b) The linear theory matter power spectrum (solid line) and the recovered power spectrum (vertical dashed lines) at $z_{2}$. c) The linear matter PS and the recovered PS from measurement at $z_{3}$. } \label{fig5}
\end{figure}

We investigate the recovery of the real-space matter power spectrum from the redshift-space multipoles ratios in this section. From the fiducial model, one already obtains the bias factors for the different values of redshifts. One also obtains the galaxy power spectrum from multipoles by using Eq.(\ref{Pdeltadelta}). From these, one can recover the matter power spectrum. We adopt the fiducial cosmological parameters for the predictions of the linear matter power spectra. We also apply the corresponding values of the growth factors at three different redshifts. These are shown in the Fig.\ref{fig5}. In the left panel of the figure, the solid line is the linear matter power spectrum for the fiducial model and the vertical lines are the values of the recovered power spectrum with 1-$\sigma$ accuracy. The recovered matter power spectrum is consistent with the linear theory prediction. The comparisons between the linear theory power spectra and the recovered ones for $z_{2}$ and $z_{3}$ are shown in the middle and the right panel of Fig.\ref{fig5}, respectively.

\section{Conculsions}
\label{Con}
We provide the analytic formulae for the multipoles and their ratio when one includes the Finger-of-God effects in the redshift-space galaxy power spectrum. One can recover the real-space linear power spectrum from the observed values of multipoles. Especially, if one uses the ratios of the redshift-space galaxy power spectrum multipoles, then one can recover the real-space linear power spectrum with less systematic effect on the nonlinear effect on the FoG factor. Therefore, we provide the recovering method based on the ratios of multipoles. After that we investigate the consistency of current DR12 multipoles data  by comparing the measured multipole ratios with the linear theory predictions. The measured values of ratios of the quadrupoles to the monopoles are consistent with the linear theory prediction at least 73 \% for $z_1$ and $z_2$. The recovered matter power spectra from the redshift-space multipoles are consistent with the linear power spectra for all redshifts. In this analysis, we did not include the effect of irregular $\mu$ distribution on the power spectra. One might further lower the $\chi^2$ by including this effect which expected to be small though. 
 
\section{Acknowledgements}
SL is supported by Basic Science Research Program through the National Research Foundation of Korea (NRF)
funded by the Ministry of Science, ICT and Future Planning (Grant No. NRF-2015R1A2A2A01004532) and (NRF-2017R1A2B4011168).
SL also thanks the department of physics at Sungkyunkwan University for their hospitality and support during the completion of this work. The author also appreciates anonymous referee for her (his) comments and suggestions to improve our manuscript. 
 
%
\appendix
\section*{Appendix}
\label{Appendix}
In this appendix, we provide the analytic forms of multipoles and the their ratios for the different FoG factors. We also provide the covariance matrices and $\chi^2$-values of $\R_{2}^{\DR12}$ and $\R_{4}^{\DR12}$ for the different redshifts.  
\section{Lorenztian FoG} 
\setcounter{equation}{0}
\renewcommand{\theequation}{A.\arabic{equation}}
Multipoles of the linear power spectrum with the Lorentzian FoG factor given in Eq.(\ref{FoGLor}) are given by
\ba \fr{\P_{0}^{(\lin)}}{\P_{\delta \delta }^{(\lin)}} &=& \frac{f x \Bigl(6 x^2+f \left(-3+x^2\right)\Bigr)+3 \left(f-x^2\right)^2 \text{ArcTan}[x]}{3 x^5} \, , \label{P0linLor} \\
\fr{\P_{2}^{(\lin)}}{\P_{\delta \delta }^{(\lin)}} &=& \frac{\left(-90 f x^3+45 x^5+f^2 x \left(45+4 x^4\right)-15 \left(f-x^2\right)^2 \left(3+x^2\right) \text{ArcTan}[x]\right) }{6 x^7} \, , \label{P2linLor} \\
\fr{\P_{4}^{(\lin)}}{\P_{\delta \delta }^{(\lin)}} 
 &=&\frac{3 \left(f-x^2\right)^2 \Bigl(-5 x \left(21+11 x^2\right)+3 \left(35+30 x^2+3 x^4\right) \text{ArcTan}[x]\Bigr) }{8 x^9} \label{P4linLor} \, ,  \ea
 where $\text{ArcTan[x]}$ is the inverse tangent function of x. From Eqs.(\ref{P0linLor})-(\ref{P4linLor}), one obtain the multipoles ratios $\P_{2}/\P_{0}$, $\P_{4}/\P_{0}$
 
 \ba \fr{\P_{2}^{(\lin)}}{\P_{0}^{(\lin)}} &=& \frac{-90 f x^3+45 x^5+f^2 x \left(45+4 x^4\right)-15 \left(f-x^2\right)^2 \left(3+x^2\right) \text{ArcTan}[x]}{2 x^2 \Bigl(f x \left(6 x^2+f \left(-3+x^2\right)\right)+3 \left(f-x^2\right)^2 \text{ArcTan}[x]\Bigr)} \, , \label{P2oP0Lor} \\
\fr{\P_{4}^{(\lin)}}{\P_{0}^{(\lin)}} 
&=& \frac{9 \left(f-x^2\right)^2 \left(-5 x \left(21+11 x^2\right)+3 \left(35+30 x^2+3 x^4\right) \text{ArcTan}[x]\right)}{8 x^4 \Bigl(f x \left(6 x^2+f \left(-3+x^2\right)\right)+3 \left(f-x^2\right)^2 \text{ArcTan}[x]\Bigr)} \, . \label{P4oP0Lor} \ea

\section{Conventional Lorentzian FoG}
\setcounter{equation}{0}
\renewcommand{\theequation}{B.\arabic{equation}}

From the conventional Lorentzian FoG factor given in Eq.(\ref{FoGLorh}), multipoles of the linear power spectrum are obtained as
\ba \fr{\P_{0}^{(\lin)}}{\P_{\delta \delta }^{(\lin)} } &=& \frac{2fx\Bigl(6x^2+f\left(-6+x^2\right)\Bigr)+6c\left(2f-x^2\right)^2\text{ArcTan}[c x]}{3x^5} \, , \label{P0linLorh} \\
\fr{\P_{2}^{(\lin)}}{\P_{\delta \delta}^{(\lin)}}  &=& \frac{180fx\left(f-x^2\right)+x^5\left(45+4f^2\right)-15 c \left(6+x^2\right) \left(-2 f+x^2\right)^2 \text{ArcTan}[c x]}{3x^7} \, , \label{P2linLorh} \\
\fr{\P_{4}^{(\lin)}}{\P_{\delta \delta}^{(\lin)}} 
 &=& \frac{-15 x\left(-2 f+x^2\right)^2 \left(42+11 x^2\right)+9 c \left(-2 f+x^2\right)^2 \left(140+60 x^2+3 x^4\right) \text{ArcTan}[c x]}{4 x^9} \, , \label{P2linLorh}\ea
where  $c = \sqrt{0.5}$. From Eqs.(\ref{P0linLor})-(\ref{P4linLor}), one obtains the multipoles ratios $\P_{2}/\P_{0}$, $\P_{4}/\P_{0}$
 
 \ba \fr{\P_{2}^{(\lin)}}{\P_{0}^{(\lin)}} &=& \frac{-180 f x^3+45 x^5+4 f^2 x \left(45+x^4\right)-15 c \left(6+x^2\right) \left(-2 f+x^2\right)^2 \text{ArcTan}[c x]}{2 x^2 \left(f x \left(6 x^2+f \left(-6+x^2\right)\right)+3 c \left(-2 f+x^2\right)^2 \text{ArcTan}[c x]\right)} \, , \label{P2oP0Lorh} \\
\fr{\P_{4}^{(\lin)}}{\P_{0}^{(\lin)}} 
&=& \frac{9 \left(-2 f+x^2\right)^2 \left(-5 x \left(42+11 x^2\right)+3 c \left(140+60 x^2+3 x^4\right) \text{ArcTan}[c x]\right)}{8 x^4 \left(f x \left(6 x^2+f \left(-6+x^2\right)\right)+3 c \left(-2 f+x^2\right)^2 \text{ArcTan}[c x]\right)} \, . \label{P4oP0Lorh} \ea

\section{Covariance matrices and $\chi^2$ } 
\setcounter{equation}{0}
\renewcommand{\theequation}{C.\arabic{equation}}

\begin{table} 
\centering
\begin{tabular}{l c ccc}
\hline  \\[0.1ex]
                  &  & $z_{1} = 0.38$ & $z_{2} = 0.51$  & $z_{3} = 0.61$ \\[1ex] \hline \\[0.1ex]
\multirow{2}{*}{$\chi^2$} & $\R_2$  &  6.10 & 2.16 &  7.90 \\[1ex] \cdashline{2-5} \\[0.1ex]
                  & $\R_4$ &131.56 & 34.78 & 22.38  \\[1ex] \hline \\[0.1ex]
                  & d.o.f & 9 & 8 & 8 \\[1ex] \hline
\end{tabular}
\caption{$\chi^2$-values and probabilities of $\R_2$ and $\R_{4}$ at three redshifts and the degrees of freedom (d.o.f).}
\label{tab-1}
\end{table}

We determine the covariance matrices of $\R_{2}$ and $\R_{4}$ between the different k-modes at the three redshifts. We also obtain the $\chi^2$ values. These are shown in the table.\ref{tab-1}.
Covariance matrices for $\R_{2}$ and $\R_{4}$ at $z_{1} = 0.38$ are given by.
\be
C_{ij}^{\R_2} (z_1) =
\tiny{\begin{pmatrix}
1.93\times 10^{-1}&8.92\times 10^{-1}&4.99\times 10^{-2}&3.79\times 10^{-2}&-3.26\times 10^{-3}&1.53\times 10^{-2}&-1.39\times 10^{-2}&-9.8\times 10^{-3}&-2.82\times 10^{-2}\\
8.92\times 10^{-1}&7.19\times 10^1&1.1&8.61\times 10^{-1}&4.95\times 10^{-1}&5.2\times 10^{-1}&3.11\times 10^{-1}&2.77\times 10^{-1}&1.68\times 10^{-1}\\
4.99\times 10^{-2}&1.1&8.7\times 10^{-2}&6.93\times 10^{-2}&4.46\times 10^{-2}&4.43\times 10^{-2}&3.07\times 10^{-2}&2.72\times 10^{-2}&2.04\times 10^{-2}\\
3.79\times 10^{-2}&8.61\times 10^{-1}&6.93\times 10^{-2}&5.53\times 10^{-2}&3.6\times 10^{-2}&3.55\times 10^{-2}&2.49\times 10^{-2}&2.2\times 10^{-2}&1.68\times 10^{-2}\\
-3.26\times 10^{-3}&4.95\times 10^{-1}&4.46\times 10^{-2}&3.6\times 10^{-2}&2.81\times 10^{-2}&2.46\times 10^{-2}&2.14\times 10^{-2}&1.86\times 10^{-2}&1.77\times 10^{-2}\\
1.53\times 10^{-2}&5.2\times 10^{-1}&4.43\times 10^{-2}&3.55\times 10^{-2}&2.46\times 10^{-2}&2.33\times 10^{-2}&1.78\times 10^{-2}&1.56\times 10^{-2}&1.31\times 10^{-2}\\
-1.39\times 10^{-2}&3.11\times 10^{-1}&3.07\times 10^{-2}&2.49\times 10^{-2}&2.14\times 10^{-2}&1.78\times 10^{-2}&1.71\times 10^{-2}&1.47\times 10^{-2}&1.52\times 10^{-2}\\
-9.8\times 10^{-3}&2.77\times 10^{-1}&2.72\times 10^{-2}&2.2\times 10^{-2}&1.86\times 10^{-2}&1.56\times 10^{-2}&1.47\times 10^{-2}&1.27\times 10^{-2}&1.29\times 10^{-2}\\
-2.82\times 10^{-2}&1.68\times 10^{-1}&2.04\times 10^{-2}&1.68\times 10^{-2}&1.77\times 10^{-2}&1.31\times 10^{-2}&1.52\times 10^{-2}&1.29\times 10^{-2}&1.5\times 10^{-2} 
\end{pmatrix}} \, , 
\label{CR2z1}
\ee

\be
C_{ij}^{\R_4} (z_1) = 
\tiny{\begin{pmatrix}
5.42\times 10^{-4}&2.69\times 10^{-2}&2.89\times 10^{-3}&5.56\times 10^{-3}&-1.11\times 10^{-3}&1.66\times 10^{-3}&1.28\times 10^{-3}&2.65\times 10^{-4}&-2.03\times 10^{-3}\\
2.69\times 10^{-2}&3.29\times 10^1&3.32\times 10^{-1}&4.71\times 10^{-1}&1.02\times 10^{-2}&1.71\times 10^{-1}&1.36\times 10^{-1}&6.22\times 10^{-2}&-8.14\times 10^{-2}\\
2.89\times 10^{-3}&3.32\times 10^{-1}&2.5\times 10^{-2}&3.88\times 10^{-2}&-1.27\times 10^{-3}&1.38\times 10^{-2}&1.11\times 10^{-2}&4.67\times 10^{-3}&-8.53\times 10^{-3}\\
5.56\times 10^{-3}&4.71\times 10^{-1}&3.88\times 10^{-2}&6.59\times 10^{-2}&-7.19\times 10^{-3}&2.17\times 10^{-2}&1.71\times 10^{-2}&5.71\times 10^{-3}&-1.89\times 10^{-2}\\
-1.11\times 10^{-3}&1.02\times 10^{-2}&-1.27\times 10^{-3}&-7.19\times 10^{-3}&4.82\times 10^{-3}&-8.93\times 10^{-4}&-4.74\times 10^{-4}&1.2\times 10^{-3}&5.58\times 10^{-3}\\
1.66\times 10^{-3}&1.71\times 10^{-1}&1.38\times 10^{-2}&2.17\times 10^{-2}&-8.93\times 10^{-4}&7.72\times 10^{-3}&6.17\times 10^{-3}&2.56\times 10^{-3}&-4.94\times 10^{-3}\\
1.28\times 10^{-3}&1.36\times 10^{-1}&1.11\times 10^{-2}&1.71\times 10^{-2}&-4.74\times 10^{-4}&6.17\times 10^{-3}&4.95\times 10^{-3}&2.13\times 10^{-3}&-3.69\times 10^{-3} \\
2.65\times 10^{-4}&6.22\times 10^{-2}&4.67\times 10^{-3}&5.71\times 10^{-3}&1.2\times 10^{-3}&2.56\times 10^{-3}&2.13\times 10^{-3}&1.33\times 10^{-3}&-4.87\times 10^{-5}\\
-2.03\times 10^{-3}&-8.14\times 10^{-2}&-8.53\times 10^{-3}&-1.89\times 10^{-2}&5.58\times 10^{-3}&-4.94\times 10^{-3}&-3.69\times 10^{-3}&-4.87\times 10^{-5}&8.49\times 10^{-3}
\end{pmatrix}} \, .
\label{CR4z1}
\ee

Covariance matrices for $\R_{2}$ and $\R_{4}$ at $z_{2} = 0.51$ are denoted by $C_{ij}^{\R_2}(z_2)$ and $C_{ij}^{\R_4}(z_2)$. For $k = 0.015$ h/Mpc, $\R_{2}$ and $\R_{4}$ values are undetermined and thus the analysis is done except this mode. The number of degrees of freedom (d.o.f) at this redshift is 8.
\be
C_{ij}^{\R_2} (z_2) =
\scriptsize{\begin{pmatrix}
2.85\times 10^{-1}&6.33\times 10^{-2}&1.76\times 10^{-2}&1.71\times 10^{-2}&2.97\times 10^{-3}&2.95\times 10^{-3}&1.5\times 10^{-2}&-9.73\times 10^{-3} \\
6.33\times 10^{-2}&8.19\times 10^{-2}&5.38\times 10^{-2}&4.44\times 10^{-2}&3.31\times 10^{-2}&3.04\times 10^{-2}&3.03\times 10^{-2}&2.21\times 10^{-2} \\
1.76\times 10^{-2}&5.38\times 10^{-2}&3.96\times 10^{-2}&3.23\times 10^{-2}&2.59\times 10^{-2}&2.37\times 10^{-2}&2.16\times 10^{-2}&1.93\times 10^{-2}\\
1.71\times 10^{-2}&4.44\times 10^{-2}&3.23\times 10^{-2}&2.64\times 10^{-2}&2.1\times 10^{-2}&1.93\times 10^{-2}&1.77\times 10^{-2}&1.56\times 10^{-2} \\
2.97\times 10^{-3}&3.31\times 10^{-2}&2.59\times 10^{-2}&2.1\times 10^{-2}&1.74\times 10^{-2}&1.59\times 10^{-2}&1.39\times 10^{-2}&1.36\times 10^{-2}\\
2.95\times 10^{-3}&3.04\times 10^{-2}&2.37\times 10^{-2}&1.93\times 10^{-2}&1.59\times 10^{-2}&1.47\times 10^{-2}&1.28\times 10^{-2}&1.25\times 10^{-2} \\
1.5\times 10^{-2}&3.03\times 10^{-2}&2.16\times 10^{-2}&1.77\times 10^{-2}&1.39\times 10^{-2}&1.28\times 10^{-2}&1.21\times 10^{-2}&1.02\times 10^{-2}\\
-9.73\times 10^{-3}&2.21\times 10^{-2}&1.93\times 10^{-2}&1.56\times 10^{-2}&1.36\times 10^{-2}&1.25\times 10^{-2}&1.02\times 10^{-2}&1.16\times 10^{-2}
\end{pmatrix}} \, , 
\label{CR2z2}
\ee

\be
C_{ij}^{\R_4} (z_2) =
\scriptsize{\begin{pmatrix}
6.64\times 10^{-3}&7.57\times 10^{-3}&4.59\times 10^{-3}&-2.63\times 10^{-3}&6.78\times 10^{-5}&1.56\times 10^{-3}&-2.36\times 10^{-4}&-3.58\times 10^{-3} \\
7.57\times 10^{-3}&4.43\times 10^{-2}&2.59\times 10^{-2}&-1.75\times 10^{-2}&-9.52\times 10^{-4}&8.1\times 10^{-3}&-2.5\times 10^{-3}&-2.24\times 10^{-2}\\
4.59\times 10^{-3}&2.59\times 10^{-2}&1.54\times 10^{-2}&-8.77\times 10^{-3}&1.97\times 10^{-4}&5.12\times 10^{-3}&-7.99\times 10^{-4}&-1.18\times 10^{-2}\\
-2.63\times 10^{-3}&-1.75\times 10^{-2}&-8.77\times 10^{-3}&1.59\times 10^{-2}&4.9\times 10^{-3}&-9.97\times 10^{-4}&4.97\times 10^{-3}&1.68\times 10^{-2}\\
6.78\times 10^{-5}&-9.52\times 10^{-4}&1.97\times 10^{-4}&4.9\times 10^{-3}&2.32\times 10^{-3}&9.61\times 10^{-4}&2.08\times 10^{-3}&4.52\times 10^{-3} \\
1.56\times 10^{-3}&8.1\times 10^{-3}&5.12\times 10^{-3}&-9.97\times 10^{-4}&9.61\times 10^{-4}&2.06\times 10^{-3}&5.43\times 10^{-4}&-2.14\times 10^{-3}\\
-2.36\times 10^{-4}&-2.5\times 10^{-3}&-7.99\times 10^{-4}&4.97\times 10^{-3}&2.08\times 10^{-3}&5.43\times 10^{-4}&1.93\times 10^{-3}&4.82\times 10^{-3} \\
-3.58\times 10^{-3}&-2.24\times 10^{-2}&-1.18\times 10^{-2}&1.68\times 10^{-2}&4.52\times 10^{-3}&-2.14\times 10^{-3}&4.82\times 10^{-3}&1.85\times 10^{-2}\end{pmatrix}} \, . 
\label{CR4z2}
\ee

Covariance matrices for $\R_{2}$ and $\R_{4}$ at $z_{3} = 0.61$ are given by.
\be
C_{ij}^{\R_2} (z_3) =
\scriptsize{\begin{pmatrix}
1.82\times 10^{-1}&6.08\times 10^{-2}&1.53\times 10^{-2}&2.57\times 10^{-2}&2.37\times 10^{-2}&6.58\times 10^{-3}&2.06\times 10^{-2}&1.26\times 10^{-2} \\
6.08\times 10^{-2}&7.18\times 10^{-2}&4.28\times 10^{-2}&3.89\times 10^{-2}&3.39\times 10^{-2}&2.44\times 10^{-2}&2.84\times 10^{-2}&2.28\times 10^{-2} \\
1.53\times 10^{-2}&4.28\times 10^{-2}&4.1\times 10^{-2}&3.32\times 10^{-2}&2.85\times 10^{-2}&2.44\times 10^{-2}&2.36\times 10^{-2}&2.04\times 10^{-2}\\
2.57\times 10^{-2}&3.89\times 10^{-2}&3.32\times 10^{-2}&2.8\times 10^{-2}&2.42\times 10^{-2}&1.95\times 10^{-2}&2.02\times 10^{-2}&1.69\times 10^{-2}\\
2.37\times 10^{-2}&3.39\times 10^{-2}&2.85\times 10^{-2}&2.42\times 10^{-2}&2.09\times 10^{-2}&1.68\times 10^{-2}&1.75\times 10^{-2}&1.47\times 10^{-2}\\
6.58\times 10^{-3}&2.44\times 10^{-2}&2.44\times 10^{-2}&1.95\times 10^{-2}&1.68\times 10^{-2}&1.46\times 10^{-2}&1.39\times 10^{-2}&1.21\times 10^{-2}\\
2.06\times 10^{-2}&2.84\times 10^{-2}&2.36\times 10^{-2}&2.02\times 10^{-2}&1.75\times 10^{-2}&1.39\times 10^{-2}&1.46\times 10^{-2}&1.23\times 10^{-2}\\
1.26\times 10^{-2}&2.28\times 10^{-2}&2.04\times 10^{-2}&1.69\times 10^{-2}&1.47\times 10^{-2}&1.21\times 10^{-2}&1.23\times 10^{-2}&1.05\times 10^{-2}
\end{pmatrix}} \, , 
\label{CR2z3}
\ee

\be
C_{ij}^{\R_4} (z_3) =
\scriptsize{\begin{pmatrix}
2.19\times 10^{-3}&-2.39\times 10^{-4}&-1.1\times 10^{-3}&6.27\times 10^{-4}&2.2\times 10^{-4}&1.43\times 10^{-3}&-1.85\times 10^{-3}&-8.19\times 10^{-4}\\
-2.39\times 10^{-4}&3.67\times 10^{-3}&4.11\times 10^{-3}&2.65\times 10^{-3}&2.38\times 10^{-3}&1.64\times 10^{-3}&2.79\times 10^{-3}&2.11\times 10^{-3}\\
-1.1\times 10^{-3}&4.11\times 10^{-3}&1.1\times 10^{-2}&1.87\times 10^{-3}&3.09\times 10^{-3}&-2.92\times 10^{-3}&1.2\times 10^{-2}&6.81\times 10^{-3}\\
6.27\times 10^{-4}&2.65\times 10^{-3}&1.87\times 10^{-3}&2.61\times 10^{-3}&1.98\times 10^{-3}&2.71\times 10^{-3}&9.75\times 10^{-5}&6.86\times 10^{-4}\\
2.2\times 10^{-4}&2.38\times 10^{-3}&3.09\times 10^{-3}&1.98\times 10^{-3}&1.79\times 10^{-3}&1.22\times 10^{-3}&2.13\times 10^{-3}&1.61\times 10^{-3} \\
1.43\times 10^{-3}&1.64\times 10^{-3}&-2.92\times 10^{-3}&2.71\times 10^{-3}&1.22\times 10^{-3}&5.26\times 10^{-3}&-5.91\times 10^{-3}&-2.47\times 10^{-3} \\
-1.85\times 10^{-3}&2.79\times 10^{-3}&1.2\times 10^{-2}&9.75\times 10^{-5}&2.13\times 10^{-3}&-5.91\times 10^{-3}&1.48\times 10^{-2}&7.84\times 10^{-3} \\
-8.19\times 10^{-4}&2.11\times 10^{-3}&6.81\times 10^{-3}&6.86\times 10^{-4}&1.61\times 10^{-3}&-2.47\times 10^{-3}&7.84\times 10^{-3}&4.33\times 10^{-3}
\end{pmatrix}} \, . 
\label{CR4z3}
\ee


\begin{thebibliography}{99}
 
 \bibitem{160703150} F.~Beutler, {\it et. al.}, Mon.\ Not.\ R.\ Astr.\ Soc. {\bf 466}, 2242 (2017). [arXiv:1607.03150].

\bibitem{TNS} A.~Taruya, T.~Nishimichi, and S.~Saito, Phys.\ Rev.\ D {\bf 82}, 063522 (2010) [arXiv:astro-ph/1006.0699].

\bibitem{0407214} R.~Scoccimarro, Phys.\ Rev.\ D {\bf 70}, 083007 (2004) [arXiv:astro-ph/0407214].
 
\bibitem{Jackson} J.~C.~Jackson, Mon.\ Not.\ R.\ Astr.\ Soc. {\bf 156}, 1 (1972).  
 
\bibitem{9603031} J.~A.~Peacock and S.~J.~Dodds, Mon.\ Not.\ R.\ Astr.\ Soc. {\bf 280}, L19 (1996) [arXiv:astro-ph/9603031]. 
 
\bibitem{Kaiser} N.~Kaiser, Mon.\ Not.\ R.\ Astr.\ Soc. {\bf 227}, 1 (1987). 
 
\bibitem{AP} C.~Alcock and B.~Paczynski, Nature {\bf 281}, 358 (1979). 
 
\bibitem{9605017} W.~E.~Ballinger, J.~A.~Peacock, and A.~F.~Heavens,  Mon.\ Not.\ R.\ Astr.\ Soc. {\bf 282}, 877 (1996) [arXiv:astro-ph/9605017]. 

\bibitem{12036594} L.~Anderson, {\it et. al.}, Mon.\ Not.\ R.\ Astr.\ Soc. {\bf 427}, 3435 (2012) [arXiv:1203.6594]. 

\bibitem{12065309} L.~Samushia, {\it et. al.},  Mon.\ Not.\ R.\ Astr.\ Soc. {\bf 429}, 1514 (2013) [arXiv:1206.5309]. 

\bibitem{160703153}  G.-B.~Zhao, {\it et. al.}, Mon.\ Not.\ R.\ Astr.\ Soc. {\bf 466}, 762 (2016) [arXiv:1607.03153]. 
 
\bibitem{0409207} K.~Yamamoto, B.~A.~Bassett, and H.~Nishioka, Phys.\ Rev.\ Lett. {\bf 94}, 051301 (2005) [arXiv:astro-ph/0409207].

\bibitem{13124611} F.~Beutler, {\it et.al.}, Mon.\ Not.\ Roy.\ Astron.\ Soc. {\bf 443}, 1065 (2014) [arXiv:1312.4611].

\bibitem{150201589} P.~A.~R.~Ade, {\it et.al.} [Planck Collaboration], [arXiv:1502.01589].

\bibitem{11054165} B.~A.~Reid and M.~White, Mon.\ Not.\ R.\ Astr.\ Soc. {\bf 417}, 1913 (2011) [arXiv:1105.4165]. 

\bibitem{Peacock92} J.~A.~Peacock, Mon.\ Not.\ R.\ Astr.\ Soc. {\bf 258}, 581 (1992)

\bibitem{DavisPeebles} M.~Davis and P.~J.~E.~Peebles,  Astro.\ Phys.\ J {\bf 267}, 465 (1983).

\bibitem{14077325} S.~Lee, C.~Park, and S.~G.~Biern, Phys.\ Lett.\ B {\bf 736}, 403 (2014) [arXiv:1407.7325].


\bibitem{0111575} M.~Tegmark, A.~J.~S.~Hamilton, and Y.~Xu, Mon.\ Not.\ Roy.\ Astron.\ Soc. {\bf 335}, 887 (2002) [arXiv:astro-ph/0111575]. 

\bibitem{0310725} M.~Tegmark and {\it et.al.}, Astro.\ Phys.\ J {\bf 606}, 702 (2004) [arXiv:astro-ph/0310725].




\end{thebibliography}
\end{document}